\documentclass[10pt,twocolumn,twoside]{IEEEtran}
\usepackage{amssymb}
\usepackage{graphicx}
\usepackage{amsmath}
\usepackage{array}
\newcommand{\PreserveBackslash}[1]{\let\temp=\\#1\let\\=\temp}
\newcolumntype{C}[1]{>{\PreserveBackslash\centering}p{#1}}
\newcolumntype{R}[1]{>{\PreserveBackslash\raggedleft}p{#1}}
\newcolumntype{L}[1]{>{\PreserveBackslash\raggedright}p{#1}}
\usepackage[usenames]{color}
\usepackage{colortbl,booktabs}
\usepackage{tabularx}
\usepackage{multicol}
\usepackage{booktabs}
\usepackage[Symbol]{upgreek}
\usepackage{subfigure}
\usepackage{bm}
\usepackage{cite}
\usepackage{balance}
\usepackage{threeparttable}
\usepackage{amsmath}
\usepackage{dcolumn}
\usepackage{multirow}
\usepackage{booktabs}
\usepackage{amsfonts}
\newcolumntype{d}[1]{D{.}{.}{#1}}
\usepackage{algorithm} 
\usepackage{algorithmic} 
\usepackage{amsmath}

\usepackage{enumerate}
\usepackage{balance}

\begin{document}

\bibliographystyle{IEEEtran} 

\title{Spectrum and Energy Efficient Beamspace MIMO-NOMA for Millimeter-Wave Communications Using Lens Antenna Array}

\author{\IEEEauthorblockN{Bichai Wang, Linglong Dai, Zhaocheng Wang, Ning Ge, and Shidong Zhou}
\thanks{Manuscript received January 27, 2017; revised May 15, 2017; accepted May 23, 2017. Date of publication XX, 2017; date of current version XX, 2017.}
\thanks{This work was supported by the National Key Basic Research Program of China (Grant No. 2013CB329203), the National Natural Science Foundation of China (Grant Nos. 61571267 and 61571270), and the Royal Academy of Engineering under the UK-China Industry Academia Partnership Programme Scheme (Grant No. UK-CIAPP$\backslash$49).}
\thanks{All authors are with the Tsinghua National Laboratory for Information Science and Technology (TNList) as well as the Department of Electronic Engineering, Tsinghua University, Beijing 100084, P. R. China (E-mail: wbc15@mails.tsinghua.edu.cn, \{daill, zcwang, gening, zhousd\}@tsinghua.edu.cn).} %

}

\maketitle
\begin{abstract}

The recent concept of beamspace multiple input multiple output
(MIMO) can significantly reduce the number of required
radio-frequency (RF) chains in millimeter-wave (mmWave) massive MIMO
systems without obvious performance loss. However, the fundamental limit of existing
beamspace MIMO is that, the number of supported users cannot be
larger than the number of RF chains at the same time-frequency resources. To break this fundamental limit, in this paper we propose a new spectrum and
energy efficient mmWave transmission scheme that integrates the
concept of non-orthogonal multiple access (NOMA) with beamspace
MIMO, i.e., beamspace MIMO-NOMA. By using NOMA in beamspace MIMO
systems, the number of supported users can be larger than the number
of RF chains at the same time-frequency resources. Particularly,
the achievable sum rate of the proposed beamspace MIMO-NOMA
in a typical mmWave channel model is analyzed, which shows an obvious performance gain compared with the existing beamspace MIMO. Then, a precoding scheme based on the principle
of zero-forcing (ZF) is designed to reduce the inter-beam
interferences in the beamspace MIMO-NOMA system. Furthermore, to maximize the achievable sum rate, a dynamic
power allocation is proposed by solving the joint power optimization
problem, which not only includes the intra-beam power optimization,
but also considers the inter-beam power optimization. Finally, an
iterative optimization algorithm with low complexity is developed to realize the dynamic power
allocation. Simulation results show that the proposed beamspace
MIMO-NOMA can achieve higher spectrum and energy efficiency compared with existing beamspace MIMO.

\end{abstract}


\section{Introduction}\label{S1}

\IEEEPARstart WITH the rapid development of the Mobile Internet and
the Internet of Things (IoT), challenging requirements for the 5th
generation (5G) of wireless communication systems are expected to be
satisfied, which are fuelled by the prediction that the global mobile
data traffic will grow in the range of 10-100 times from 2020 to
2030. The emerging millimeter-wave (mmWave)
communications, operating from 30-300 GHz, provide an opportunity to
meet such explosive capacity demand for 5G~\cite{MMWAVE}.
In addition to orders-of-magnitude lager bandwidths, the smaller
wavelengths at mmWave allow more antennas in a
same physical space, which enables massive multiple input
multiple output (MIMO) to provide more multiplexing gain and beamforming gain~\cite{JBBS,HX1,ASBSMIMO,PARF,XGEE,HX2,XC1}. In fact, it has been
demonstrated that mmWave massive MIMO can
achieve orders-of-magnitude increase in system capacity~\cite{XGEE}.

However, it is difficult to
realize mmWave massive MIMO in practice due to high
transceiver complexity and energy
consumption~\cite{PARF}~\cite{XGBS}. Particularly, each antenna in
MIMO systems usually requires one dedicated radio-frequency (RF)
chain~\cite{FRMIMO}. Therefore, the use of a very large number of
antennas in mmWave massive MIMO systems leads to an equally large
number of RF chains. Moreover, it is shown that RF components may
consume up to $70 \%$ of the total transceiver energy
consumption~\cite{PARF}~\cite{SCEE}. As a result, the hardware cost
and energy consumption caused by a large number of RF chains in mmWave massive MIMO systems become
unaffordable in practice.

To address this challenging problem, a lot of
studies have been done to reduce the hardware complexity and energy
consumption. Particularly, the antenna selection technique has been
considered to solve this problem~\cite{SSAS,AMAS,SSCAS}. However, an obvious performance
loss will be introduced. Recently, the concept of beamspace
MIMO has been proposed in the pioneering
work~\cite{JBBS} to significantly reduce the number of required RF chains in mmWave massive MIMO systems. By
using the lens antenna array, which plays a role in realizing spatial discrete Fourier transformation~\cite{YZMM}, beamspace MIMO can transform the
conventional spatial channel to the beamspace channel to capture the
channel sparsity at mmWave frequencies~\cite{XGBS}. Accordingly, the dominant beams are selected according to the sparse beamspace channel to
reduce the number of required RF chains. Moreover, by the
use of lens antenna array, narrow beams can be preserved even
with a reduced number of RF chains, which allows to significantly
reduce the power required per beam and the inter-beam
interferences~\cite{PARF}. Therefore, unlike the antenna selection
technique, the performance of beamspace MIMO with beam selection is
close-to-optimal~\cite{ASBSMIMO}~\cite{PARF}. Nevertheless, a fundamental limit of beamspace MIMO that was explicitly or implicitly considered in all published papers
on beamspace MIMO~\cite{PARF,XGBS,JBBS,YZMM,ASBSMIMO} is that, each RF chain can only
support one user at the same time-frequency resources, so the maximum number of users that can be
supported cannot exceed the number of RF
chains. The reason is that, the degree of freedom (DoF) provided by the RF chains must be larger than or equal to the DoF required by users, otherwise signal for different users cannot be separated by linear operation.

In this paper, we aim to break this fundamental limit by proposing a spectrum and energy efficient mmWave
transmission scheme that integrates the new concept of non-orthogonal
multiple access (NOMA) with beamspace MIMO, i.e., beamspace
MIMO-NOMA\footnote{Simulation codes are provided to reproduce the
results presented in this paper:
http://oa.ee.tsinghua.edu.cn/dailinglong/publications/publications.html.}. Particularly, NOMA has
also been considered as a promising candidate for 5G to improve
spectrum efficiency and connectivity
density~\cite{ZD1,KH2015nonsic,ZD2,SMRNOMA,ZDAMIMO,dai2015non,ZDLDMIMONOMA,ZDMMIMO}.
In contrast to the orthogonal multiple access (OMA) schemes relying on the time-,
frequency-, code-domain or on their combinations, NOMA can be
realized in a new domain, i.e., the power domain. By performing
superposition coding at the transmitter and successive interference
cancellation (SIC) at the receiver, multiple users can be
simultaneously supported at the same time-frequency-space resources,
and the channel gain difference among users can be translated into
multiplexing gain by superposition coding. By integrating NOMA into
beamspace MIMO, potential performance gain can be achieved. Specifically, The contributions of
this paper can be summarized as follows.

\begin{enumerate}

\item We propose a new spectrum and energy efficient mmWave transmission scheme, i.e., beamspace MIMO-NOMA, that combines the advantages of NOMA and beamspace MIMO. To the best of our knowledge, this is the first work using NOMA as a potential multiple access scheme for beamspace MIMO in mmWave communications. Particularly, by using intra-beam superposition coding and SIC, more than one user can be simultaneously supported in one beam, which is essentially different from existing beamspace MIMO using one beam to only serve one user. Thus, the number of supported users can be larger than the number of RF chains at the same time-frequency resources in the proposed beamspace MIMO-NOMA scheme, and the achievable sum rate in a typical mmWave channel model can be also significantly improved. Note that although the combination of spatial MIMO and NOMA has been widely investigated~\cite{YSNOMA,YHPA,BKNOMABF,MSADPA,[25],QSNOMA,QZRBF,[28]}, it only focused on the conventional MIMO systems rather than the mmWave massive MIMO systems. Therefore, the existing MIMO-NOMA schemes~\cite{YSNOMA,YHPA,BKNOMABF,MSADPA,[25],QSNOMA,QZRBF,[28]} have not considered the transmission characteristics in mmWave communications, e.g., the channel sparsity, as well as the uncertainty of the number of conflicting users in beamspace MIMO systems.

\item To reduce the inter-beam interferences in the proposed beamspace MIMO-NOMA system, the equivalent channel vector is determined for each beam to realize precoding based on the principle of zero-forcing (ZF). On the one hand, when the line-of-sight (LoS) component of users' channels is dominant, the high correlation~\cite{[29]} of users' beamspace channels in the same beam at mmWave frequencies is utilized to generate the equivalent channel vectors. Note that potential performance gain can be achieved by exploiting the high channel correlation in NOMA~\cite{ZDLDMIMONOMA}. On the other hand, when the LoS component does not exist or the effect of non-line-of-sight (NLoS) components is significant, the channel correlation in the same beam may be not high enough. To this end, we also consider the singular value decomposition (SVD)-based equivalent channel, which exploits all of the beamspace channel vectors of users in the same beam.

\item In the proposed beamspace MIMO-NOMA scheme, in addition to intra-beam interferences caused by superposition coding,
users also suffer from interferences from other beams. Thus, a direct combination of NOMA and beamspace MIMO is not able to guarantee the reliable performance
in practice. In most existing MIMO-NOMA schemes,
the fixed inter-beam power allocation is usually exploited without
optimization, and intra-beam power optimization is considered only
for two users. On the contrary, in the proposed beamspace MIMO-NOMA
scheme, a dynamic power allocation scheme is realized to maximize the achievable sum rate
with the transmitted power constraint by solving the
joint power optimization problem, which not only includes the
intra-beam power optimization, but also considers the inter-beam
power optimization. Furthermore, an iterative optimization algorithm
is developed to realize the dynamic power allocation, and the convergence as
well as computational complexity of this
algorithm are also analyzed.

\item We verify the performance of
the proposed beamspace MIMO-NOMA scheme by simulations. The convergence of the developed iterative
optimization algorithm for dynamic power allocation is validated, and it is shown that only 10 times of iteration are required to make it converged.
Furthermore, we show that the proposed beamspace MIMO-NOMA can achieve higher
spectrum and energy efficiency than that of existing beamspace MIMO
systems, e.g., $25 \%$ energy efficiency gain can be achieved.

\end{enumerate}

The rest of this paper is organized as follows. The system model of
the proposed beamspace MIMO-NOMA system is introduced in
Section II. Section III analyzes the achievable sum rate of the
proposed beamspace MIMO-NOMA system, and Section IV introduces the
precoding scheme based on the principle of ZF. In Section V, a dynamic
power allocation scheme is proposed to maximize the achievable sum rate.
Simulation results are provided in Section VI. Finally, conclusions
are drawn in Section VII.

\emph{Notation}: We use upper-case and lower-case boldface letters
to denote matrices and vectors, respectively; $(\cdot)^T$,
$(\cdot)^H$, $(\cdot)^{-1}$, $(\cdot)^{\dagger}$, tr($\cdot$), and
$\|\cdot\|_p$ denote the transpose, conjugate transpose, matrix
inversion, Moore-Penrose matrix inversion, the trace of the matrix,
and $l_p$ norm operation, respectively. ${\rm diag}\{ {\bf{p}}\}$ denotes the diagonal matrix whose diagonal elements consist of the elements in the vector ${\bf{p}}$. ${\rm E}\left\{  \cdot  \right\}$ denotes the expectation. $|\Gamma|$ denotes the
number of elements in set $\Gamma$. ${\bf{A}}{\left( {i,:} \right)_{i \in \Gamma }}$ denotes the submatrix of $\bf{A}$ that consists of the $i$th row of $\bf{A}$ for all ${i \in \Gamma }$. We use the notation ${\cal{CN}}\left( \bf{m}, \bf{R} \right)$ to denote the complex Gaussian distribution with mean $\bf{m}$ and covariance $\bf{R}$. Finally, $\mathbf{I}_N$ is the
$N \times N$ identity matrix.

\section{System Model}\label{S2}

In this paper, we consider a single-cell downlink mmWave communication system, where the base
station (BS) is equipped with $N$ antennas and $N_{\rm RF}$ RF
chains, and $K$ single-antenna users
are simultaneously served by the BS~\cite{ASBSMIMO}~\cite{PARF}. The system
model of existing beamspace MIMO will be introduced at first in this section, and
then the proposed beamspace MIMO-NOMA scheme will be presented in detail.

\subsection{Beamspace MIMO}

In traditional MIMO systems as shown in Fig. 1 (a), the received
signal vector ${\bf{y}} = {\left[ {{y_1},{y_2}, \cdots ,{y_K}}
\right]^T}$ can be expressed as
\begin{equation}\label{eq4}
{\bf{y}} = {{\bf{H}}^H}{\bf{WPs}} + {\bf{v}},
\end{equation}
where ${\bf{s}} = {\left[ {{s_1},{s_2}, \cdots ,{s_K}} \right]^T}$
is the $K \times 1$ transmitted signal vector for all $K$ users with
normalized power ${\rm E}\left( {{\bf{s}}{{\bf{s}}^H}} \right) =
{{\bf{I}}_K}$, ${\bf{P}} = {\rm diag}\{ {\bf{p}}\} $ includes the
transmitted power for all $K$ users where ${\bf{p}} = \left[ {\sqrt
{{p_1}} ,\sqrt {{p_2}} , \cdots ,\sqrt {{p_K}} } \right]$ satisfies
$\sum\limits_{k = 1}^K {{p_k}}  \le P$ (the maximum
transmitted power at the BS), ${\bf{W}} = \left[
{{{\bf{w}}_1},{{\bf{w}}_2}, \cdots ,{{\bf{w}}_K}} \right]$ is the $N
\times K$ precoding matrix with ${\left\| {{{\bf{w}}_k}}
\right\|_2} = 1$ for $k = 1,2, \cdots ,K$, and ${\bf{v}}$ is the noise
vector following the distribution ${\cal{CN}}\left( 0,{\sigma
^2}{{\bf{I}}_{K}} \right)$. Finally, ${\bf{H}} = \left[
{{{\bf{h}}_1},{{\bf{h}}_2}, \cdots ,{{\bf{h}}_K}} \right]$ of size
$N \times K$ is the channel matrix, where ${\bf{h}}_k$ of size $N \times 1$ denotes the spatial channel vector between the BS and the $k$th user. Particularly, in this paper, we consider the widely used Saleh-Valenzuela channel
model for mmWave communications~\cite{PARF,XGBS,JBBS,YZMM,ASBSMIMO}, so
${{\bf{h}}_k}$ can be represented as
\begin{equation}\label{eq1}
{{\bf{h}}_k} = \beta _k^{(0)}{\bf{a}}\left( {\theta  _k^{(0)}} \right) + \sum\limits_{l = 1}^L {\beta _k^{\left( l \right)}{\bf{a}}\left( {\theta _k^{\left( l \right)}} \right)},
\end{equation}
where $\beta _k^{(0)}{\bf{a}}( {\theta _k^{(0)}} )$ is
the LoS component of the $k$th user, in which $\beta
_k^{(0)}$ denotes the complex gain and ${\bf{a}}( {\theta
_k^{(0)}} )$ represents the spatial direction. ${\beta
_k^{\left( l \right)}{\bf{a}}( {\theta _k^{\left( l \right)}}
)}$ for $1 \le l \le L$ is the $l$th NLoS
component of the $k$th user, where $L$ is the total number of NLoS
components. ${\bf{a}}\left( \theta  \right)$ is the $N \times 1$
array steering vector. Note that at mmWave frequencies, the
amplitudes $\{ {| {\beta _k^{\left( l \right)}} |}
\}_{l = 1}^N$ of NLoS components are typically 5 to 10 dB
weaker than the amplitude $| {\beta _k^{(0)}} |$ of the
LoS component~\cite{ASBSMIMO}~\cite{TSRICC}.

For the typical uniform linear array (ULA)~\cite{PARF}, the array steering vector ${\bf{a}}\left( \theta  \right)$ can
be expressed as
\begin{equation}\label{eq2}
{\bf{a}}\left( \theta  \right) = \frac{1}{{\sqrt N }}{\left[ {{e^{ - j2\pi \theta m}}} \right]_{m \in J\left( N \right)}},
\end{equation}
where $J\left( N \right) = \left\{ {i - \left( {N - 1} \right)/2,i =
0,1, \cdots ,N - 1} \right\}$ is a symmetric set of indices centered
around zero. The spatial direction is defined as $\theta  =
\frac{d}{\lambda }\sin \left( \phi  \right)$, where $\phi$ is the
physical direction satisfying $ - \frac{\pi }{2} \le \phi  \le
\frac{\pi }{2}$, $\lambda$ is the signal wavelength, and $d$ is the
antenna spacing.

\begin{figure}[tp]
\begin{center}
\includegraphics[width=1\linewidth]{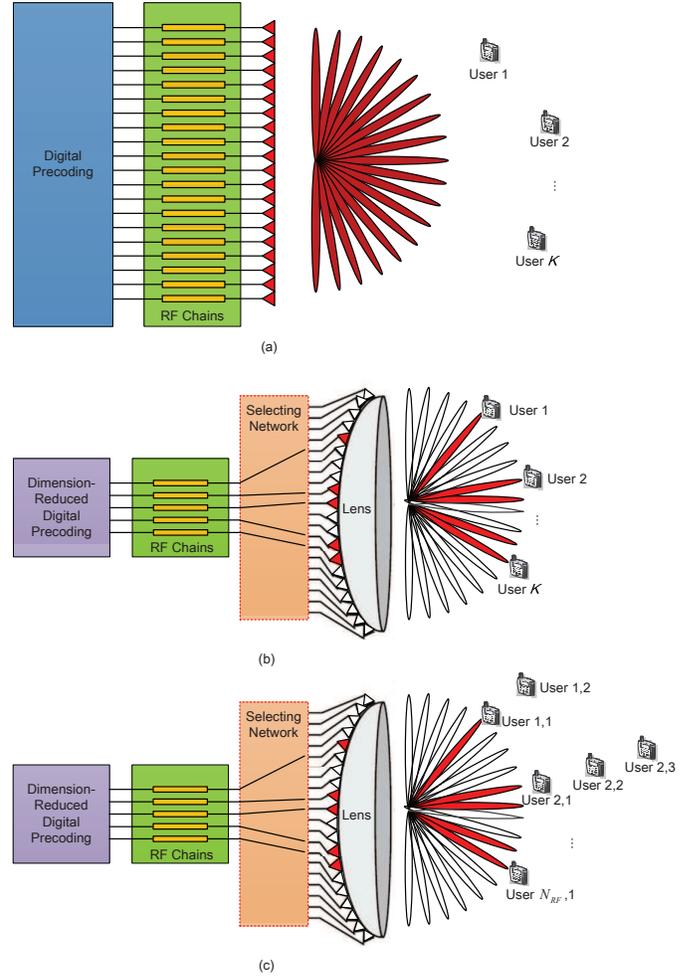} \caption{System models of MIMO architectures: (a) traditional MIMO; (b) beamspace MIMO; (c) the proposed beamspace MIMO-NOMA.}
\end{center}
\end{figure}

As shown in Fig. 1 (a), in traditional MIMO systems, the number of required RF chains is equal to the number of BS antennas, i.e., ${N_{\rm RF}} = N$, which is usually large for mmWave massive MIMO systems, e.g., ${N_{\rm RF}} = N = 256$~\cite{XGBS}~\cite{SHLS}. Therefore, the direct application of massive MIMO at mmWave frequencies is prohibitive due to high hardware cost and energy consumption caused by RF chains~\cite{PARF}~\cite{SCEE}, e.g., about 250 mW is consumed by each RF chain, and 64 W is required by
a mmWave massive MIMO system with 256 antennas~\cite{XGEE}.

To address this issue, the concept of beamspace MIMO has been recently proposed, which can utilize lens antenna array to significantly reduce the number of required
RF chains in mmWave massive MIMO systems without
obvious performance loss. As shown in Fig. 1. (b), by employing lens antenna array, the channel (\ref{eq1}) in the spatial domain can be transformed to the beamspace channel~\cite{YZMM}. Specifically, the mathematical function of the lens antenna array is to realize the spatial discrete Fourier transformation with the $N \times N$ transform matrix $\bf{U}$~\cite{XGBS}, which contains the array steering vectors of $N$ directions covering the entire space as follows:
\begin{equation}\label{eq5}
{\bf{U}} = {\left[ {{\bf{a}}\left( {{{\bar \theta }_1}} \right),{\bf{a}}\left( {{{\bar \theta }_2}} \right), \cdots ,{\bf{a}}\left( {{{\bar \theta }_N}} \right)} \right]^H},
\end{equation}
where ${{\bar \theta }_n} = \frac{1}{N}\left( {n - \frac{{N +
1}}{2}} \right)$ for $n = 1,2, \cdots ,N$ are the predefined spatial
directions. Then, the received signal vector $\bf{\bar y}$ in beamspace MIMO
systems can be represented as
\begin{equation}\label{eq6}
\begin{array}{l}
{\bf{\bar y}}  = {{\bf{H}}^H}{{\bf{U}}^H}{\bf{WPs}} + {\bf{v}}
     = {{{\bf{\bar H}}}^H}{\bf{WPs}} + {\bf{v}},
\end{array}
\end{equation}
where the beamspace channel matrix ${{\bf{\bar H}}}$ is defined as
\begin{equation}\label{eq7}
{\bf{\bar H}}
      = {\bf{UH}}
      = \left[ {{\bf{U}}{{\bf{h}}_1},{\bf{U}}{{\bf{h}}_2}, \cdots ,{\bf{U}}{{\bf{h}}_K}} \right]
      = \left[ {{{{\bf{\bar h}}}_1},{{{\bf{\bar h}}}_2}, \cdots ,{{{\bf{\bar h}}}_K}} \right],
\end{equation}
where ${{{{\bf{\bar h}}}_k}}={\bf{U}}{{\bf{h}}_k}$ is the beamspace channel vector between the BS and the $k$th user, which is the Fourier transformation of the spatial channel vector ${\bf{h}}_k$ in (\ref{eq1}).

As for the beamspace channel matrix ${{\bf{\bar H}}}$ defined in
(\ref{eq7}), each row of ${{\bf{\bar H}}}$ corresponds to one beam,
and all $N$ rows correspond to $N$ beams with spatial
directions ${{\bar \theta }_1},{{\bar \theta }_2}, \cdots ,{{\bar
\theta }_N}$, separately. In mmWave communications, since the number
of dominant scatters is very limited, the number of NLoS components
$L$ is much smaller than the number of beams
$N$~\cite{XGBS}. Therefore, the number of dominant elements of each
beamspace channel vector ${{{{\bf{\bar h}}}_k}}$ is much smaller
than $N$, namely, the beamspace channel matrix ${{\bf{\bar H}}}$ has
a sparse nature~\cite{YZMM}. This sparse structure can be exploited
to design dimension-reduced beamspace MIMO systems without obvious
performance loss by beam selection~\cite{ASBSMIMO}~\cite{PARF}. Specifically, according to the
sparse beamspace channel matrix, only a small number of beams can be
selected to simultaneously serve $K$ users. Then, the received signal
vector in (\ref{eq6}) can be rewritten as
\begin{equation}\label{eq8}
{\bf{\bar y}} = {\bf{\bar H}}_r^H{{\bf{W}}_r}{\bf{Ps}} + {\bf{v}},
\end{equation}
where ${{{\bf{\bar H}}}_r} = {\bf{\bar H}}{\left( {i,:} \right)_{i
\in \Gamma }}$ of size $\left| \Gamma  \right| \times K$ is the
dimension-reduced beamspace channel matrix including selected beams,
and $\Gamma $ is the index set of selected beams. ${{\bf{W}}_r}$ of
size $\left| \Gamma  \right| \times K$ is the dimension-reduced
precoding matrix. Since the row dimension of ${{\bf{W}}_r}$ is much
smaller than $N$ (the row dimension of the original precoding matrix ${{\bf{W}}}$), the number of required RF chains can be
significantly reduced, and we have ${N_{\rm RF}} = \left| \Gamma
\right|$~\cite{XGBS}.

However, in existing beamspace MIMO systems, one beam can only
support one user at most. Therefore, the maximum number of supported
users at the same time-frequency resources is equal to the number of RF chains~\cite{XGBS}, i.e., $K \le
{N_{\rm RF}}$, which is the fundamental limit of beamspace MIMO
systems that was explicitly or implicitly considered in all published papers
on beamspace MIMO~\cite{PARF,XGBS,JBBS,YZMM,ASBSMIMO}. The reason is that, the DoF provided by the RF chains must be larger than or equal to the DoF required by users, otherwise signal for different users cannot be separated by linear operation. To break this limit, we propose a new mmWave transmission scheme that integrates NOMA with beamspace MIMO in the next subsection.

\subsection{Proposed Beamspace MIMO-NOMA}

In order to further improve spectrum efficiency and connectivity
density, we propose to leverage NOMA in beamspace mmWave massive
MIMO systems. As shown in Fig. 1
(c), unlike existing beamspace MIMO systems,
more than one user can be simultaneously served within each selected
beam in the proposed beamspace MIMO-NOMA scheme.

Specifically, beam selection algorithms, e.g., maximum magnitude
(MM) selection~\cite{ASBSMIMO} and maximization of the
signal-to-interference-plus-noise ratio (SINR)
selection~\cite{PARF}, can be used to select one beam for each user,
and each RF chain corresponds to one beam. Note that different users
are likely to select the same beam, which are called ``conflicting
users'' in this paper. Particularly, for a typical mmWave massive MIMO system with $N = 256$ antennas
and $K = 32$ users whose spatial directions follow the uniform
distribution, the probability that there exists
users selecting the same beam is $87\%$~\cite{XGBS}. In contrast to existing beamspace MIMO
systems, where user scheduling is performed to select only one user out of
these conflicting users~\cite{XGBS}, conflicting users can be simultaneously served using
the same RF chain in the proposed beamspace MIMO-NOMA system.

Although the number of selected beams is equal to $N_{\rm RF}$, the number of simultaneously served users $K$ can be larger than $N_{\rm RF}$, i.e., $K \ge {N_{\rm RF}}$. Let $S_n$ for $n = 1,2, \cdots ,{N_{\rm RF}}$ denote the set of users served by the $n$th beam with ${S_i} \cap {S_j} = \Phi $ for $i \ne j$ and $\sum\limits_{n = 1}^{{N_{\rm RF}}} {\left| {{S_n}} \right|}  = K$. The ${N_{\rm RF}} \times 1$ beamspace channel vector after beam selection between the BS and the $m$th user in the $n$th beam is denoted by ${{\bf{h}}_{m,n}}$, and ${{\bf{w}}_n}$ of size ${N_{\rm RF}} \times 1$ denotes the uniform precoding vector for users in the $n$th beam. Without loss of generality, we assume that ${\left\| {{\bf{h}}_{1,n}^H{{\bf{w}}_n}} \right\|_2} \ge {\left\| {{\bf{h}}_{2,n}^H{{\bf{w}}_n}} \right\|_2} \ge  \cdots  \ge {\left\| {{\bf{h}}_{\left| {{S_n}} \right|,n}^H{{\bf{w}}_n}} \right\|_2}$ for $n = 1,2, \cdots ,{N_{\rm RF}}$. The received signal ${y_{m,n}}$ at the $m$th user in the $n$th beam ($n = 1,2, \cdots ,{N_{\rm RF}}$, and $m = 1,2, \cdots ,\left| {{S_n}} \right|$) can be expressed as
\begin{equation}
\label{eq9}
\begin{array}{l}
\begin{aligned}
{y_{m,n}} =& {\bf{h}}_{m,n}^H\sum\limits_{j = 1}^{{N_{{\rm{RF}}}}} {\sum\limits_{i = 1}^{\left| {{S_j}} \right|} {{{\bf{w}}_j}\sqrt {{p_{i,j}}} {s_{i,j}}} }  + {v_{m,n}}\\
 =& \underbrace {{\bf{h}}_{m,n}^H{{\bf{w}}_n}\sqrt {{p_{m,n}}} {s_{m,n}}}_{\rm desired \ signal} \\
  &+ \underbrace {{\bf{h}}_{m,n}^H{{\bf{w}}_n}\sum\limits_{i = 1}^{m - 1} {\sqrt {{p_{i,n}}} {s_{i,n}}}  + {\bf{h}}_{m,n}^H{{\bf{w}}_n}\sum\limits_{i = m + 1}^{\left| {{S_n}} \right|} {\sqrt {{p_{i,n}}} {s_{i,n}}} }_{{\rm intra-beam \ interferences}} \\
 &+ \underbrace {{\bf{h}}_{m,n}^H\sum\limits_{j \ne n} {\sum\limits_{i = 1}^{\left| {{S_j}} \right|} {{{\bf{w}}_j}\sqrt {{p_{i,j}}} {s_{i,j}}} } }_{{\rm inter-beam \ interferences}} + \underbrace {{v_{m,n}}}_{\rm noise},
\end{aligned}
\end{array}
\end{equation}
where ${s_{m,n}}$ and ${{p_{m,n}}}$ are the transmitted signal and transmitted power for the $m$th user in the $n$th beam, and ${v_{m,n}}$ is the noise following the distribution ${\cal{CN}}\left( 0,{\sigma ^2} \right)$. Note that in existing beamspace MIMO systems, only one user can be supported in each selected beam, i.e., $\left| {{S_n}} \right| = 1$ ($n = 1,2, \cdots {N_{\rm RF}}$), while $\left| {{S_n}} \right|$ ($n = 1,2, \cdots {N_{\rm RF}}$) can be larger than one in the proposed beamspace MIMO-NOMA system.

In the second equation of (\ref{eq9}), the first term is the desired
signal, the second and third terms are intra-beam interferences, the
fourth term is inter-beam interference, and the last term is
the noise. Particularly, the precoding vectors $\left\{
{{{\bf{w}}_n}} \right\}_{n = 1}^{{N_{\rm RF}}}$ should be carefully
designed to restrain inter-beam
interferences, which will be discussed later in Section IV. Intra-beam interferences caused by superposition
coding in NOMA can be suppressed by carrying out SIC according to
the increasing order of equivalent channel
gains\footnote{In this paper, we assume the beamspace channel is known by the BS.
Actually, efficient tools of compressive sensing can be utilized to reliably estimate
the beamspace channel with low pilot overhead thanks to the sparsity of
beamspace channel in mmWave massive MIMO systems~\cite{JBBS}~\cite{XGBS}.}~\cite{KH2015nonsic,SMRNOMA,dai2015non}, i.e., the $m$th user
in the $n$th beam can remove the interferences from the $i$th user
(for all $i > m$) in the $n$th beam by performing SIC.

By employing NOMA in beamspace MIMO systems, more than one user can
be simultaneously served within each beam, and thus the total number
of supported users can be larger than the number of beams, i.e., $K
\ge {N_{\rm RF}}$. However, in the proposed beamspace MIMO-NOMA
system, in addition to intra-beam interferences caused by
superposition transmission, users also suffer from interferences
from other beams. Thus, a straightforward combination of NOMA and
beamspace MIMO is not able to guarantee the reliable performance in practice, and
precoding as well as power allocation should be designed to reduce
interferences by maximizing the achievable sum rate.

\section{Achievable Sum Rate}\label{S3}

As discussed in the previous section, in the $n$th beam using NOMA
with SIC, the $i$th ($i > m$) user's signal is detectable at the
$m$th user, provided that it is detectable at
itself~\cite{KH2015nonsic,SMRNOMA,dai2015non}, as the equivalent
channel gain of the $m$th user is larger than that of the $i$th
user, i.e., ${\left\| {{\bf{h}}_{1,n}^H{{\bf{w}}_n}} \right\|_2} \ge
{\left\| {{\bf{h}}_{2,n}^H{{\bf{w}}_n}} \right\|_2} \ge  \cdots  \ge
{\left\| {{\bf{h}}_{\left| {{S_n}} \right|,n}^H{{\bf{w}}_n}}
\right\|_2}$ as assumed before. Therefore, the $m$th user can detect
the $i$th user's signals for $1 \le m < i \le \left| {{S_n}}
\right|$, and then remove the detected signals from its received
signals, in a successive manner. Then, the remaining received signal
at the $m$th user in the $n$th beam can be rewritten as
\begin{equation}\label{eq10}
\begin{array}{l}
\begin{aligned}
{\hat y_{m,n}} =& \underbrace {{\bf{h}}_{m,n}^H{{\bf{w}}_n}\sqrt {{p_{m,n}}} {s_{m,n}}}_{\rm desired \ signal} + \underbrace {{\bf{h}}_{m,n}^H{{\bf{w}}_n}\sum\limits_{i = 1}^{m - 1} {\sqrt {{p_{i,n}}} {s_{i,n}}}}_{{\rm intra-beam \ interferences}} \\
&+ \underbrace {{\bf{h}}_{m,n}^H\sum\limits_{j \ne n} {\sum\limits_{i = 1}^{\left| {{S_j}} \right|} {{{\bf{w}}_j}\sqrt {{p_{i,j}}} {s_{i,j}}} } }_{{\rm inter-beam \ interferences}} + \underbrace {{v_{m,n}}}_{\rm noise}.
\end{aligned}
\end{array}
\end{equation}

Then, according to (\ref{eq10}), the SINR at the $m$th user in the $n$th beam can be presented as
\setcounter{equation}{9}
\begin{equation}\label{eq11}
{\gamma _{m,n}} = \frac{{\left\| {{\bf{h}}_{m,n}^H{{\bf{w}}_n}} \right\|_2^2{p_{m,n}}}}{{{\xi _{m,n}}}},
\end{equation}
where
\begin{equation}\label{eq12}
{\xi _{m,n}} = \left\| {{\bf{h}}_{m,n}^H{{\bf{w}}_n}} \right\|_2^2\sum\limits_{i = 1}^{m - 1} {{p_{i,n}}}  + \sum\limits_{j \ne n} {\left\| {{\bf{h}}_{m,n}^H{{\bf{w}}_j}} \right\|_2^2} \sum\limits_{i = 1}^{\left| {{S_j}} \right|} {{p_{i,j}}}  + {\sigma ^2}.
\end{equation}

As a result, the achievable rate at the $m$th user in the $n$th beam is
\begin{equation}\label{eq13}
{R_{m,n}} = {\log _2}\left( {1 + {\gamma _{m,n}}} \right).
\end{equation}
Finally, the achievable sum rate of the proposed beamspace MIMO-NOMA scheme is
\begin{equation}\label{eq14}
{R_{\rm sum}} = \sum\limits_{n = 1}^{{N_{\rm RF}}} {\sum\limits_{m =
1}^{\left| {{S_n}} \right|} {{R_{m,n}}} } ,
\end{equation}
which can be improved by carefully designing the precoding $\left\{ {{{\bf{w}}_n}} \right\}_{n = 1}^{{N_{\rm RF}}}$ and power allocation $\left\{ {{p_{m,n}}} \right\}_{m = 1,n = 1}^{\left| {{S_n}} \right|,{N_{\rm RF}}}$.

\section{Precoding in Downlink Systems}\label{S4}

In existing beamspace MIMO systems, where only one user can be
served in each beam (i.e., $K \le {N_{\rm RF}}$), the classical
linear ZF precoding with low complexity can be utilized to remove
the inter-beam interferences~\cite{PARF,XGBS,YZMM,ASBSMIMO}, which
can be simply realized by the pseudo-inverse of the beamspace
channel matrix for all users. However, in the proposed beamspace
MIMO-NOMA system, the number of users is larger than the number of
beams, i.e., $K \ge {N_{\rm RF}}$, which means the pseudo-inverse of
the beamspace channel matrix of size ${N_{\rm RF}} \times K$ does
not exist. As a result, the conventional ZF precoding cannot be directly
used.

To address this problem, an equivalent channel can be
determined for each beam to generate the precoding vector.
Specifically, we introduce two methods to generate the equivalent
channel for each beam, i.e., the strongest user-based equivalent
channel and singular value decomposition (SVD)-based equivalent
channel, which will be discussed in the following two subsections, separately.

\subsection{The strongest user-based equivalent channel}

As mentioned before, the
amplitudes of NLoS components are typically 5 to 10 dB
weaker than the amplitude of the
LoS component~\cite{ASBSMIMO}~\cite{TSRICC}. Therefore, the LoS component can primarily characterize the
multipath channel in mmWave communications~\cite{PARF}. At the same time, the
most important property of the beamspace channel matrix is that it
has the sparse structure representing the directions of different
users~\cite{ASBSMIMO}. As a result, if the LoS component exists, the sparse beamspace channel vectors of
different users in the same beam are highly correlated. That is to
say, one of the beamspace channel vectors for multiplexed users in
the $n$th beam can be regarded as the equivalent channel vector of
the $n$th beam. Particularly, considering that the first user in each
beam should perform SIC to decode all the other users' signals in
this beam, we use the beamspace channel vector of the first user in
each beam as the equivalent channel vector. Specifically, the
equivalent channel matrix of size ${N_{\rm RF}} \times {N_{\rm RF}}$
for all $N_{\rm RF}$ beams can be written as
\begin{equation}\label{eq15}
{\bf{\tilde H}} = \left[ {{{\bf{h}}_{1,1}},{{\bf{h}}_{1,2}}, \cdots
,{{\bf{h}}_{1,{N_{\rm RF}}}}} \right].
\end{equation}
Then, the precoding matrix of size ${N_{\rm RF}} \times {N_{\rm RF}}$ can be
generated by
\begin{equation}\label{eq16}
{\bf{\tilde W}} = \left[ {{{\bf{\tilde w}}_1},{{\bf{\tilde w}}_2}, \cdots ,{{\bf{\tilde w}}_{{N_{\rm RF}}}}} \right]
 = {\left( {{\bf{\tilde H}}} \right)^\dag }
 = {{{\bf{\tilde H}}}}{\left( {{\bf{\tilde H}}^H{{{\bf{\tilde H}}}}} \right)^{ - 1}}.
\end{equation}
After normalizing the precoding vectors, the precoding vector for
the $n$th beam ($n = 1,2, \cdots ,{N_{\rm RF}}$) can be written as
\begin{equation}\label{eq17}
{{\bf{w}}_n} = \frac{{{{{\bf{\tilde w}}}_n}}}{{{{\left\|
{{{{\bf{\tilde w}}}_n}} \right\|}_2}}}.
\end{equation}

In this precoding scheme, the first user in each beam can completely
remove the inter-beam interferences, i.e.,
\begin{equation}\label{eq18}
\frac{{{\bf{h}}_{1,j}^H{{\bf{w}}_n}}}{{{{\left\|
{{\bf{h}}_{1,j}^H{{\bf{w}}_n}} \right\|}_2}}} = \left\{
{\begin{array}{*{20}{c}}
{0,  \ \ \   {\rm for} \ j \ne n},\\
{1,  \ \ \   {\rm for} \ j = n},
\end{array}} \right.
\end{equation}
where $1 \le j,n \le {N_{\rm RF}}$. Thus, after performing SIC, the SINR
at the first user in the $n$th beam can be rewritten as
\begin{equation}\label{eq19}
{\gamma _{1,n}} = \frac{{\left\| {{\bf{h}}_{1,n}^H{{\bf{w}}_n}}
\right\|_2^2{p_{1,n}}}}{{{\sigma ^2}}}.
\end{equation}

\subsection{SVD-based equivalent channel}

When the LoS component does not exist or the effect of NLoS components is significant, the channel correlation in the same beam may be not high enough. Therefore, we also consider another precoding scheme that exploits all of the
beamspace channel vectors of users in the same beam, which is
inspired by the precoding scheme used in conventional MU-MIMO systems~\cite{QHZF}. Specifically,
let ${{\bf{H}}_n}$ of size ${N_{\rm RF}} \times \left| {{S_n}} \right|$
denote the beamspace channel matrix of all $\left| {{S_n}} \right|$ users in the $n$th beam, i.e.,
\begin{equation}\label{eq20}
{{\bf{H}}_n} = \left[ {{{\bf{h}}_{1,n}},{{\bf{h}}_{2,n}}, \cdots
,{{\bf{h}}_{\left| {{S_n}} \right|,n}}} \right],
\end{equation}
where $1 \le n \le {N_{\rm RF}}$. Then, by taking the SVD of
${\bf{H}}_n^T$, we have
\begin{equation}\label{eq21}
{\bf{H}}_n^T = {{\bf{U}}_n}{{\bf{\Sigma }}_n}{\bf{V}}_n^H,
\end{equation}
where ${{\bf{U}}_n}$ is the left singular matrix of size $\left|
{{S_n}} \right| \times \left| {{S_n}} \right|$, ${{\bf{\Sigma }}_n}$
is the $\left| {{S_n}} \right| \times {N_{\rm RF}}$ singular value
matrix with its diagonal entries sorted in a non-increasing order,
and ${{\bf{V}}_n}$ is the right singular matrix of size ${N_{\rm RF}}
\times {N_{\rm RF}}$. Then, the equivalent channel vector of the $n$th beam
can be generated by
\begin{equation}\label{eq22}
{{{\bf{\tilde h}}}_n} = {{\bf{H}}_n}{\bf{u}}_n^*,
\end{equation}
where ${\bf{u}}_n$ is the first column of ${{\bf{U}}_n}$, i.e., the
left singular vector corresponds to the maximum singular value. Finally, the
equivalent channel matrix of size ${N_{\rm RF}} \times {N_{\rm RF}}$ can be expressed as
\begin{equation}\label{eq23}
{\bf{\tilde H}} = \left[ {{{{\bf{\tilde h}}}_1},{{{\bf{\tilde h}}}_2}, \cdots ,{{{\bf{\tilde h}}}_{{N_{{\rm{RF}}}}}}} \right]\\
 = \left[ {{{\bf{H}}_1}{\bf{u}}_1^*,{{\bf{H}}_2}{\bf{u}}_2^*, \cdots ,{{\bf{H}}_{{N_{\rm RF}}}}{\bf{u}}_{{N_{\rm RF}}}^*} \right].
\end{equation}

Similar to the strongest user-based equivalent channel, the
ZF precoding matrix of size ${N_{\rm RF}} \times {N_{\rm RF}}$ can be generated
according to (\ref{eq16}) and (\ref{eq17}).

Note that in addition to ZF, other classical precoding schemes,
e.g., minimum mean square error (MMSE) and Wiener filter
(WF)~\cite{ASBSMIMO}, are also feasible based on the equivalent
channel matrix in (\ref{eq15}) or (\ref{eq23}).

After obtaining the precoding vectors, power allocation for different users in different beams will be optimized in the next section to maximize the achievable sum rate (\ref{eq14}) of the proposed beamspace MIMO-NOMA scheme.

\section{Dynamic Power Allocation}\label{S5}

The channel gain difference among users can be translated into multiplexing gains by superposition coding in NOMA systems. Therefore, power allocation has an important effect on the system performance. In fact, to suppress inter-user interferences and improve the achievable sum rate, lots of studies have been done to design power allocation in existing MIMO-NOMA systems. The integration of NOMA and MIMO was investigated in~\cite{YSNOMA}, where two users were considered in each beam with a random beamforming, and fixed power allocation schemes were utilized at the BS. In addition, fixed power allocation strategies have also been considered in~\cite{KH2015nonsic}. A coordinated frequency block-dependent inter-beam power allocation was proposed in~\cite{YHPA} to generate distinct power levels for different beams. The authors in~\cite{BKNOMABF} considered equal power allocation for different groups, and intra-group power allocation has been optimized to maximize the achievable sum rate, where each group only included two single-antenna users. The intra-group power optimization has also been investigated in~\cite{MSADPA}~\cite{[25]}, where a convex optimization algorithm was utilized to obtain the closed-form solution to power allocation. In~\cite{QSNOMA}, a non-convex power allocation problem was formulated for MIMO-NOMA systems, where only two
users have been considered, and sub-optimal solutions were provided. The authors in~\cite{QZRBF} investigated joint optimization of beamforming and power allocation, where channel uncertainties have been considered to maximize the worst-case achievable sum rate. The performance with only two users in each group was evaluated in the simulations.

Similar to existing MIMO-NOMA works, both inter-beam interferences and intra-beam interferences should be reduced to improve the achievable sum rate of the proposed beamspace MIMO-NOMA system. However, in contrast to existing MIMO-NOMA works, where fixed inter-beam power allocation and fixed number of users in each beam (e.g., two users in each beam) are usually considered, multiple users (e.g., 1, 2, or 3 users) are allowed in each beam in the proposed beamspace MIMO-NOMA scheme. Accordingly, a dynamic power allocation scheme to maximize the achievable sum rate is proposed by solving the joint power optimization problem, which not only includes the intra-beam power optimization, but also considers the inter-beam power optimization. The power
allocation problem can be formulated as
\begin{equation}\label{eq26}
\begin{array}{l}
\begin{aligned}
\mathop {\max }\limits_{\left\{ {{p_{m,n}}} \right\}} & \sum\limits_{n = 1}^{{N_{\rm RF}}} {\sum\limits_{m = 1}^{\left| {{S_n}} \right|} {{R_{m,n}}} } \\
{\rm s.t.} \ \ \ & {C_1}: \ {p_{m,n}} \ge 0, \ \ \forall n,m, \\
& {C_2}: \ \sum\limits_{n = 1}^{{N_{\rm RF}}} {\sum\limits_{m = 1}^{\left| {{S_n}} \right|} {{p_{m,n}}} }  \le P,\\
& {C_3}: \ {R_{m,n}} \ge {R_{\min }}, \ \ \forall n,m,
\end{aligned}
\end{array}
\end{equation}
where ${{R_{m,n}}}$ is the achievable rate of the $m$the user in the
$n$th beam as defined in (\ref{eq13}), the constraint $C_1$
indicates that the power allocated to each user must be positive,
$C_2$ is the transmitted power constraint with $P$ being the maximum
total transmitted power by the BS, and $C_3$ is the data
rate constraint for each user with ${R_{\min }}$ being the minimum
data rate for each user. By substituting
(\ref{eq11})-(\ref{eq13}) into the constraint $C_3$ in (\ref{eq26}),
we have
\begin{equation}\label{eq27}
\begin{array}{l}
\begin{aligned}
&\left\| {{\bf{h}}_{m,n}^H{{\bf{w}}_n}} \right\|_2^2{p_{m,n}} - \eta \left\| {{\bf{h}}_{m,n}^H{{\bf{w}}_n}} \right\|_2^2\sum\limits_{i = 1}^{m - 1} {{p_{i,n}}}
 \\
 &- \eta \sum\limits_{j \ne n} {\left\| {{\bf{h}}_{m,n}^H{{\bf{w}}_j}} \right\|_2^2} \sum\limits_{i = 1}^{\left| {{S_j}} \right|} {{p_{i,j}}}  \ge \omega,
\end{aligned}
\end{array}
\end{equation}
where $\eta  = {2^{{R_{\min }}}} - 1$ and $\omega  = \eta {\sigma
^2}$. In this way, the non-linear constraint $C_3$ has been transformed into linear constraint. Then, the optimization problem (\ref{eq26}) can be rewritten as
\begin{equation}\label{eq28}
\begin{array}{l}
\mathop {\max }\limits_{\left\{ {{p_{m,n}}} \right\}} \sum\limits_{n = 1}^{{N_{\rm RF}}} {\sum\limits_{m = 1}^{\left| {{S_n}} \right|} {{\log _2}\left( {1 +{\gamma _{m,n}}} \right)} } \\
{\rm s.t.} \ \ \ {C_1}: \ {p_{m,n}} \ge 0, \ \ \forall n,m, \\
\ \ \ \ \ \ \ {C_2}: \ \sum\limits_{n = 1}^{{N_{\rm RF}}} {\sum\limits_{m = 1}^{\left| {{S_n}} \right|} {{p_{m,n}}} }  \le P,\\
\ \ \ \ \ \ \ {C_3}: \ \begin{aligned} &\left\| {{\bf{h}}_{m,n}^H{{\bf{w}}_n}} \right\|_2^2{p_{m,n}} - \eta \left\| {{\bf{h}}_{m,n}^H{{\bf{w}}_n}} \right\|_2^2\sum\limits_{i = 1}^{m - 1} {{p_{i,n}}} \\
&- \eta \sum\limits_{j \ne n} {\left\|
{{\bf{h}}_{m,n}^H{{\bf{w}}_j}} \right\|_2^2} \sum\limits_{i =
1}^{\left| {{S_j}} \right|} {{p_{i,j}}}  \ge \omega, \ \ \forall n,m.
\end{aligned}
\end{array}
\end{equation}
We can see from (\ref{eq28}) that all constraints, i.e.,
$C_1$, $C_2$, and $C_3$, are linear inequality constraints, while
the objective function is non-convex. Therefore, this optimization problem is NP-hard, and it is very difficult to
obtain the closed-form solution to the optimal power allocation problem
(\ref{eq28}).

To solve this difficult non-convex problem (\ref{eq28}), we propose an iterative optimization
algorithm to realize power allocation. Specifically, according to the extension of the
Sherman-Morrison-Woodbury formula~\cite{QZRBF}~\cite{JRMDC}, i.e.,
\begin{equation}\label{eq29}
{\left( {{\bf{A}} + {\bf{BCD}}} \right)^{ - 1}} = {{\bf{A}}^{ - 1}} - {{\bf{A}}^{ - 1}}{\bf{B}}{\left( {{\bf{I}} + {\bf{CD}}{{\bf{A}}^{ - 1}}{\bf{B}}} \right)^{ - 1}}{\bf{CD}}{{\bf{A}}^{ - 1}},
\end{equation}
we have
\begin{equation}\label{eq30}
\begin{array}{l}
\begin{aligned}
&{\left( {1 + {\gamma _{m,n}}} \right)^{ - 1}} \\
 =& 1 - \left\| {{\bf{h}}_{m,n}^H{{\bf{w}}_n}} \right\|_2^2{p_{m,n}}{\left( {\left\| {{\bf{h}}_{m,n}^H{{\bf{w}}_n}} \right\|_2^2{p_{m,n}} + {\xi _{m,n}}} \right)^{ - 1}},
\end{aligned}
\end{array}
\end{equation}
where $n = 1,2, \cdots ,{N_{\rm RF}}$ and $m = 1,2, \cdots ,\left| {{S_n}} \right|$.

We can find that the expression (\ref{eq30}) has the same form as
the MMSE. Specifically, if MMSE detection is used to solve
${s_{m,n}}$ from ${{\hat y}_{m,n}}$ in (\ref{eq10}), this detection
problem can be formulated as
\begin{equation}\label{eq31}
c_{m,n}^o = \arg \mathop {\min }\limits_{{c_{m,n}}} {e_{m,n}},
\end{equation}
where
\begin{equation}\label{eq32}
{e_{m,n}} = {\rm{E}}\left\{ {{{\left| {{s_{m,n}} - {c_{m,n}}{{\hat y}_{m,n}}} \right|}^2}} \right\}
\end{equation}
is the mean square error (MSE), ${c_{m,n}}$ is the channel
equalization coefficient, and $c_{m,n}^o$ is the optimal value of
${c_{m,n}}$ to minimize the MSE. Substituting (\ref{eq10}) into
(\ref{eq32}), we have
\begin{equation}\label{eq33}
\begin{array}{l}
\begin{aligned}
{e_{m,n}} =& 1 - 2{\mathop{\rm Re}\nolimits} \left( {{c_{m,n}}\sqrt {{p_{m,n}}} {\bf{h}}_{m,n}^H{{\bf{w}}_n}} \right)\\
 &+ {\left| {{c_{m,n}}} \right|^2}\left( {{p_{m,n}}\left\| {{\bf{h}}_{m,n}^H{{\bf{w}}_n}} \right\|_2^2 + {\xi _{m,n}}} \right)\\
 =& {\left| {1 - {c_{m,n}}\sqrt {{p_{m,n}}} {\bf{h}}_{m,n}^H{{\bf{w}}_n}} \right|^2}\\
 &+ {\left| {{c_{m,n}}} \right|^2}\left\| {{\bf{h}}_{m,n}^H{{\bf{w}}_n}} \right\|_2^2\sum\limits_{i = 1}^{m - 1} {{p_{i,n}}} \\
 &+ {\left| {{c_{m,n}}} \right|^2}\sum\limits_{j \ne n} {\left\| {{\bf{h}}_{m,n}^H{{\bf{w}}_j}} \right\|_2^2} \sum\limits_{i = 1}^{\left| {{S_j}} \right|} {{p_{i,j}}}
 + {\left| {{c_{m,n}}} \right|^2}{\sigma ^2}.
\end{aligned}
\end{array}
\end{equation}
Then, by solving (\ref{eq31}) based on (\ref{eq33}), the optimal equalization coefficient $c_{m,n}^o$ can be calculated
by
\begin{equation}\label{eq34}
\begin{array}{l}
{\left. {\frac{{\partial {e_{m,n}}}}{{\partial {c_{m,n}}}}} \right|_{c_{m,n}^o}} = 0\\
 \Rightarrow  - \sqrt {{p_{m,n}}} {\bf{h}}_{m,n}^H{{\bf{w}}_n} \\
 \ \ \ \ + {\left( {c_{m,n}^o} \right)^*}\left( {{p_{m,n}}\left\| {{\bf{h}}_{m,n}^H{{\bf{w}}_n}} \right\|_2^2 + {\xi _{m,n}}} \right) = 0\\
 \Rightarrow c_{m,n}^o = {\left( {\sqrt {{p_{m,n}}} {\bf{h}}_{m,n}^H{{\bf{w}}_n}} \right)^*}{\left( {{p_{m,n}}\left\| {{\bf{h}}_{m,n}^H{{\bf{w}}_n}} \right\|_2^2 + {\xi _{m,n}}} \right)^{ - 1}}.
\end{array}
\end{equation}
Substituting (\ref{eq34}) into (\ref{eq33}), we obtain the MMSE as
\begin{equation}\label{eq35}
\begin{array}{l}
\begin{aligned}
e_{m,n}^o = & 1 - 2{\rm{Re}}\left( {c_{m,n}^o\sqrt {{p_{m,n}}} {\bf{h}}_{m,n}^H{{\bf{w}}_n}} \right) \\
&+ {\left| {c_{m,n}^o} \right|^2}\left( {{p_{m,n}}\left\| {{\bf{h}}_{m,n}^H{{\bf{w}}_n}} \right\|_2^2 + {\xi _{m,n}}} \right) \\
 = & 1 - 2{\rm{Re}}\left( {{{\left( {\sqrt {{p_{m,n}}} {\bf{h}}_{m,n}^H{{\bf{w}}_n}} \right)}^*}{{\left( {{p_{m,n}}\left\| {{\bf{h}}_{m,n}^H{{\bf{w}}_n}} \right\|_2^2} \right.}}}\right.\\
 & \left. {{{ \left. { + {\xi _{m,n}}} \right)}^{ - 1}}\sqrt {{p_{m,n}}} {\bf{h}}_{m,n}^H{{\bf{w}}_n}} \right) \\
 &+ {\left| {{{\left( {\sqrt {{p_{m,n}}} {\bf{h}}_{m,n}^H{{\bf{w}}_n}} \right)}^*}{{\left( {{p_{m,n}}\left\| {{\bf{h}}_{m,n}^H{{\bf{w}}_n}} \right\|_2^2 }\right.}}}\right.}\\
 &{\left.{{{\left.{ + {\xi _{m,n}}} \right)}^{ - 1}}} \right|^2}\left( {{p_{m,n}}\left\| {{\bf{h}}_{m,n}^H{{\bf{w}}_n}} \right\|_2^2 + {\xi _{m,n}}} \right) \\
 = & 1 - 2{p_{m,n}}\left\| {{\bf{h}}_{m,n}^H{{\bf{w}}_n}} \right\|_2^2{\left( {{p_{m,n}}\left\| {{\bf{h}}_{m,n}^H{{\bf{w}}_n}} \right\|_2^2 + {\xi _{m,n}}} \right)^{ - 1}} \\
 &+ {p_{m,n}}\left\| {{\bf{h}}_{m,n}^H{{\bf{w}}_n}} \right\|_2^2{\left( {{p_{m,n}}\left\| {{\bf{h}}_{m,n}^H{{\bf{w}}_n}} \right\|_2^2 + {\xi _{m,n}}} \right)^{ - 1}} \\
 = & 1 - {p_{m,n}}\left\| {{\bf{h}}_{m,n}^H{{\bf{w}}_n}} \right\|_2^2{\left( {{p_{m,n}}\left\| {{\bf{h}}_{m,n}^H{{\bf{w}}_n}} \right\|_2^2 + {\xi _{m,n}}} \right)^{ - 1}}.
\end{aligned}
\end{array}
\end{equation}
which is equal to ${\left( {1 + {\gamma _{m,n}}} \right)^{ - 1}}$ in (\ref{eq30}), i.e., we have
\begin{equation}\label{eq36}
{\left( {1 + {\gamma _{m,n}}} \right)^{ - 1}} = \mathop {\min }\limits_{{c_{m,n}}} {e_{m,n}}.
\end{equation}
Then, the achievable rate of the $m$th user in the $n$th beam can be written as
\begin{equation}\label{eq37}
\begin{array}{l}
{R_{m,n}} = {\log _2}\left( {1 + {\gamma _{m,n}}} \right)
 = \mathop {\max }\limits_{{c_{m,n}}} \left( { - {{\log }_2}{e_{m,n}}} \right).
\end{array}
\end{equation}

To remove the log function in (\ref{eq37}), we introduce
the following proposition~\cite{QZRBF}.

\emph{Proposition 1}: Let $f\left( a \right) =  - \frac{{ab}}{{\ln 2}} + {\log _2} a  + \frac{1}{{\ln 2}}$ and $a$ be a positive real number, we have
\begin{equation}\label{eq38}
\mathop {\max }\limits_{a > 0} f\left( a \right) =  - {\log _2}b,
\end{equation}
where the optimal value of $a$ is ${a^o} = \frac{1}{b}$.

\begin{proof}
The function $f\left( a \right)$ is concave, and thus the maximum value of $f\left( a \right)$ can be obtained by solving
\begin{equation}\label{eq39}
{\left. {\frac{{\partial f\left( a \right)}}{{\partial a}}} \right|_{a = {a^o}}} = 0.
\end{equation}
Then, we have ${a^o} = \frac{1}{b}$. By instituting ${a^o}$ into $f\left( a \right)$, the maximum value of $f\left( a \right)$ is $- {\log _2}b$.
\end{proof}

Using Proposition 1, (\ref{eq37}) can be rewritten as
\begin{equation}\label{eq40}
{R_{m,n}} = \mathop {\max }\limits_{{c_{m,n}}} \mathop {\max }\limits_{{a_{m,n}} > 0} \left( { - \frac{{{a_{m,n}}{e_{m,n}}}}{{\ln 2}} + {{\log }_2}{a_{m,n}} + \frac{1}{{\ln 2}}} \right).
\end{equation}
As a result, the objective function for the optimization problem (\ref{eq28}) has been transformed into quadratic programming function, and (\ref{eq28}) can be reformulated as
\begin{equation}\label{eq41}
\begin{array}{l}
\mathop {\max }\limits_{\left\{ {{p_{m,n}}} \right\}} \sum\limits_{n = 1}^{{N_{\rm RF}}} {\sum\limits_{m = 1}^{\left| {{S_n}} \right|} \mathop {\max }\limits_{{c_{m,n}}} \mathop {\max }\limits_{{a_{m,n}} > 0} \left( { - \frac{{{a_{m,n}}{e_{m,n}}}}{{\ln 2}} + {{\log }_2}{a_{m,n}} + \frac{1}{{\ln 2}}} \right)} \\
{\rm s.t.} \ \ \ {C_1}: \ {p_{m,n}} \ge 0, \ \ \forall n,m, \\
\ \ \ \ \ \ \ {C_2}: \ \sum\limits_{n = 1}^{{N_{\rm RF}}} {\sum\limits_{m = 1}^{\left| {{S_n}} \right|} {{p_{m,n}}} }  \le P,\\
\ \ \ \ \ \ \ {C_3}: \ \begin{aligned} &\left\| {{\bf{h}}_{m,n}^H{{\bf{w}}_n}} \right\|_2^2{p_{m,n}} - \eta \left\| {{\bf{h}}_{m,n}^H{{\bf{w}}_n}} \right\|_2^2\sum\limits_{i = 1}^{m - 1} {{p_{i,n}}} \\ &- \eta \sum\limits_{j \ne n} {\left\|
{{\bf{h}}_{m,n}^H{{\bf{w}}_j}} \right\|_2^2} \sum\limits_{i =
1}^{\left| {{S_j}} \right|} {{p_{i,j}}}  \ge \omega, \ \ \forall n,m.
\end{aligned}
\end{array}
\end{equation}

To solve the reformulated optimization problem (\ref{eq41}), we propose to iteratively optimize
$\{ {{c_{m,n}}} \}$, $\{ {{a_{m,n}}} \}$, and $\{ {{p_{m,n}}} \}$.
Specifically, given the optimal power allocation solution $\left\{
{p_{m,n}^{\left( {t - 1} \right)}} \right\}$ in the $(t-1)$th
iteration, the optimal solution of $\left\{ {c_{m,n}^{\left( {t}
\right)}} \right\}$ in the $t$th iteration can be obtained according to (\ref{eq34}), i.e.,
\begin{equation}\label{eq42}
\begin{array}{l}
\begin{aligned}
&c_{m,n}^{(t)}\\
=& {\left( {\sqrt {p_{m,n}^{\left( {t - 1} \right)}}
{\bf{h}}_{m,n}^H{{\bf{w}}_n}} \right)^*}{\left( {p_{m,n}^{\left( {t
- 1} \right)}\left\| {{\bf{h}}_{m,n}^H{{\bf{w}}_n}} \right\|_2^2 +
\xi _{m,n}^{\left( {t - 1} \right)}} \right)^{ - 1}},
\end{aligned}
\end{array}
\end{equation}
where
\begin{equation}\label{eq43}
\begin{array}{l}
\begin{aligned}
\xi _{m,n}^{\left( {t - 1} \right)} =& \left\| {{\bf{h}}_{m,n}^H{{\bf{w}}_n}} \right\|_2^2\sum\limits_{i = 1}^{m - 1} {p_{i,n}^{\left( {t - 1} \right)}}\\
&+ \sum\limits_{j \ne n} {\left\| {{\bf{h}}_{m,n}^H{{\bf{w}}_j}}
\right\|_2^2} \sum\limits_{i = 1}^{\left| {{S_j}} \right|}
{p_{i,j}^{\left( {t - 1} \right)}}  + {\sigma ^2},
\end{aligned}
\end{array}
\end{equation}
and the corresponding MMSE denoted by (\ref{eq35}) in the $t$th iteration can be expressed as
\begin{equation}\label{eq44}
\begin{array}{l}
\begin{aligned}
&e_{m,n}^{o \left( t \right)}\\
 =& 1 - \left\| {{\bf{h}}_{m,n}^H{{\bf{w}}_n}} \right\|_2^2p_{m,n}^{\left( {t - 1} \right)}{\left( {\left\| {{\bf{h}}_{m,n}^H{{\bf{w}}_n}} \right\|_2^2p_{m,n}^{\left( { t- 1} \right)} + \xi _{m,n}^{\left( {t - 1} \right)}} \right)^{ - 1}}.
\end{aligned}
\end{array}
\end{equation}
Then, the optimal solution of $\left\{ {a_{m,n}^{\left( {t}
\right)}} \right\}$ in the $t$th iteration can be obtained by
\begin{equation}\label{eq45}
a_{m,n}^{\left( t \right)} = \frac{1}{{e_{m,n}^{o \left(
t\right)}}}.
\end{equation}

After obtaining the optimal $\left\{ {c_{m,n}^{\left( {t} \right)}} \right\}$
and $\left\{ {a_{m,n}^{\left( {t} \right)}} \right\}$ in the $t$th iteration, the optimal $\left\{
{p_{m,n}^{\left( {t} \right)}} \right\}$ in the $t$th iteration can
be obtained by solving the following problem:
\begin{equation}\label{eq46}
\begin{array}{l}
\mathop {\min }\limits_{\left\{ {{p_{m,n}^{(t)}}} \right\}} \sum\limits_{n = 1}^{{N_{\rm RF}}} {\sum\limits_{m = 1}^{\left| {{S_n}} \right|} a_{m,n}^{\left( t \right)}e_{m,n}^{\left( t \right)}} \\
{\rm s.t.} \ \ \ {C_1}: \ {p_{m,n}^{(t)}} \ge 0, \ \ \forall n,m, \\
\ \ \ \ \ \ \ {C_2}: \ \sum\limits_{n = 1}^{{N_{\rm RF}}} {\sum\limits_{m = 1}^{\left| {{S_n}} \right|} {{p_{m,n}^{(t)}}} }  \le P,\\
\ \ \ \ \ \ \ {C_3}: \ \begin{aligned} & \left\| {{\bf{h}}_{m,n}^H{{\bf{w}}_n}} \right\|_2^2{p_{m,n}^{(t)}} - \eta \left\| {{\bf{h}}_{m,n}^H{{\bf{w}}_n}} \right\|_2^2\sum\limits_{i = 1}^{m - 1} {{p_{i,n}^{(t)}}} \\ &- \eta \sum\limits_{j \ne n} {\left\|
{{\bf{h}}_{m,n}^H{{\bf{w}}_j}} \right\|_2^2} \sum\limits_{i =
1}^{\left| {{S_j}} \right|} {{p_{i,j}^{(t)}}}  \ge \omega, \ \
\forall n,m,
\end{aligned}
\end{array}
\end{equation}
where
\begin{equation}\label{eq47}
\begin{array}{l}
\begin{aligned}
e_{m,n}^{\left( t \right)} =& {\left| {1 - c_{m,n}^{(t)}\sqrt {p_{m,n}^{\left( t \right)}} {\bf{h}}_{m,n}^H{{\bf{w}}_n}} \right|^2}\\
 &+ {\left| {c_{m,n}^{(t)}} \right|^2}\left\| {{\bf{h}}_{m,n}^H{{\bf{w}}_n}} \right\|_2^2\sum\limits_{i = 1}^{m - 1} {p_{i,n}^{\left( t \right)}} \\
 &+ {\left| {c_{m,n}^{(t)}} \right|^2}\sum\limits_{j \ne n} {\left\| {{\bf{h}}_{m,n}^H{{\bf{w}}_j}} \right\|_2^2} \sum\limits_{i = 1}^{\left| {{S_j}} \right|} {p_{i,j}^{\left( t \right)}}
 + {\left| {c_{m,n}^{(t)}} \right|^2}{\sigma ^2}.
\end{aligned}
\end{array}
\end{equation}

To solve the convex optimization problem (\ref{eq46}), we define the Lagrange function as
\begin{equation}\label{eq48}
\begin{array}{l}
\begin{aligned}
L\left( {p,\lambda ,\mu } \right)
 =& \sum\limits_{n = 1}^{{N_{\rm RF}}} {\sum\limits_{m = 1}^{\left| {{S_n}} \right|} {a_{m,n}^{\left( t \right)}} e_{m,n}^{\left( t \right)}}
 + \lambda \left( {\sum\limits_{n = 1}^{{N_{\rm RF}}} {\sum\limits_{m = 1}^{\left| {{S_n}} \right|} {p_{m,n}^{\left( t \right)}} }  - P} \right)\\
 &+ \sum\limits_{n = 1}^{{N_{\rm RF}}} {\sum\limits_{m = 1}^{\left| {{S_n}} \right|} {{\mu _{m,n}}{\theta _{m,n}}} },
\end{aligned}
\end{array}
\end{equation}
where
\begin{equation}\label{eq47+}
\begin{array}{l}
\begin{aligned}
{\theta _{m,n}} = & \eta \left\| {{\bf{h}}_{m,n}^H{{\bf{w}}_n}} \right\|_2^2\sum\limits_{i = 1}^{m - 1} {p_{i,n}^{(t)}}  + \eta \sum\limits_{j \ne n} {\left\| {{\bf{h}}_{m,n}^H{{\bf{w}}_j}} \right\|_2^2} \sum\limits_{i = 1}^{\left| {{S_j}} \right|} {p_{i,j}^{(t)}}\\
& - \left\| {{\bf{h}}_{m,n}^H{{\bf{w}}_n}} \right\|_2^2p_{m,n}^{(t)} + \omega,
\end{aligned}
\end{array}
\end{equation}
$\lambda \ge 0$, and ${\mu _{m,n}} \ge 0$ ($n = 1,2, \cdots
,{N_{\rm RF}}$, $m = 1,2, \cdots ,\left| {{S_n}} \right|$). Then, the
Karush-Kuhn-Tucker (KKT)~\cite{SBCV} conditions of (\ref{eq46}) can be obtained by the following three equations (\ref{eq49})-(\ref{eq51}):
\begin{equation}\label{eq49}
\begin{array}{l}
\begin{aligned}
\frac{{\partial L}}{{\partial {p_{m,n}}}} =& a_{m,n}^{\left( t
\right)}\left(
{\left| {c_{m,n}^{(t)}} \right|^2}\left\| {{\bf{h}}_{m,n}^H{{\bf{w}}_n}} \right\|_2^2 \right.\\
 &\left.- {\mathop{\rm Re}\nolimits} \left( {c_{m,n}^{(t)}{\bf{h}}_{m,n}^H{{\bf{w}}_n}} \right){\left( {p_{m,n}^{\left( t \right)}} \right)^{ - \frac{1}{2}}}
\right)\\
 &+ \sum\limits_{u = m + 1}^{\left| {{S_n}} \right|} {a_{u,n}^{\left( t \right)}{{\left| {c_{u,n}^{(t)}} \right|}^2}\left\| {{\bf{h}}_{u,n}^H{{\bf{w}}_n}} \right\|_2^2}\\
 &+ \sum\limits_{v \ne n} {\sum\limits_{u = 1}^{\left| {{S_v}} \right|} {a_{u,v}^{\left( t \right)}{{\left| {c_{u,v}^{(t)}} \right|}^2}\left\| {{\bf{h}}_{u,v}^H{{\bf{w}}_n}} \right\|_2^2} }
 + \lambda \\
 &- {\mu _{m,n}}\left\| {{\bf{h}}_{m,n}^H{{\bf{w}}_n}} \right\|_2^2
 + \sum\limits_{u = m + 1}^{\left| {{S_n}} \right|} {{\mu _{u,n}}\eta \left\| {{\bf{h}}_{u,n}^H{{\bf{w}}_n}}
 \right\|_2^2}\\
 &+ \sum\limits_{v \ne n} {\sum\limits_{u = 1}^{\left| {{S_v}} \right|} {{\mu _{u,v}}\eta \left\| {{\bf{h}}_{u,v}^H{{\bf{w}}_n}} \right\|_2^2} } \\
 =& 0,
\end{aligned}
\end{array}
\end{equation}

\begin{equation}\label{eq50}
\lambda \left( {\sum\limits_{n = 1}^{{N_{\rm RF}}} {\sum\limits_{m =
1}^{\left| {{S_n}} \right|} {p_{m,n}^{\left( t \right)}} }  - P}
\right) = 0,
\end{equation}

\begin{equation}\label{eq51}
{\mu _{m,n}}{\theta _{m,n}} = 0, \ \ \ \forall n,m.
\end{equation}
From (\ref{eq49}), we can obtain the optimal solution of $p_{m,n}$
as follows:
\begin{equation}\label{eq52}
p_{m,n}^{\left( t \right)} = {\left( {\frac{{a_{m,n}^{\left( t
\right)}{\mathop{\rm Re}\nolimits} \left(
{c_{m,n}^{(t)}{\bf{h}}_{m,n}^H{{\bf{w}}_n}} \right)}}{\tau }}
\right)^2},
\end{equation}
where
\begin{equation}\label{eq53}
\begin{array}{l}
\begin{aligned}
\tau  =&  \sum\limits_{u = m}^{\left| {{S_n}} \right|} {a_{u,n}^{\left( t \right)}{{\left| {c_{u,n}^{(t)}} \right|}^2}\left\| {{\bf{h}}_{u,n}^H{{\bf{w}}_n}} \right\|_2^2}\\
 &+ \sum\limits_{v \ne n} {\sum\limits_{u = 1}^{\left| {{S_v}} \right|} {a_{u,v}^{\left( t \right)}{{\left| {c_{u,v}^{(t)}} \right|}^2}\left\| {{\bf{h}}_{u,v}^H{{\bf{w}}_n}} \right\|_2^2} }\\
 &+ \lambda
 - {\mu _{m,n}}\left\| {{\bf{h}}_{m,n}^H{{\bf{w}}_n}} \right\|_2^2
 + \sum\limits_{u = m + 1}^{\left| {{S_n}} \right|} {{\mu _{u,n}}\eta \left\| {{\bf{h}}_{u,n}^H{{\bf{w}}_n}} \right\|_2^2}\\
 &+ \sum\limits_{v \ne n} {\sum\limits_{u = 1}^{\left| {{S_v}} \right|} {{\mu _{u,v}}\eta \left\| {{\bf{h}}_{u,v}^H{{\bf{w}}_n}} \right\|_2^2} }.
\end{aligned}
\end{array}
\end{equation}

\begin{algorithm}[tp]
\caption{Proposed iterative power allocation algorithm}
\begin{algorithmic}[1]
\REQUIRE ~~\\
    Beamspace channel vectors: ${{\bf{h}}_{m,n}}$ for $\forall n,m$;\\
    Precoding vectors: ${{\bf{w}}_n}$ for $\forall n$;\\
    Noise variance: ${\sigma ^2}$;\\
    Maximum iteration times: $T_{\max}$.
\ENSURE ~~\\
    Power allocation: $p_{m,n} $ for $\forall m,n$.

\STATE $t=0$.

\WHILE {$t < T_{\max}$}

\STATE Obtain the optimal $\{ c_{m,n}^{(t)} \}$ according to (\ref{eq42});

\STATE Obtain the optimal $\{ a_{m,n}^{(t)} \}$ according to (\ref{eq45});

\STATE Obtain the optimal $\{ p_{m,n}^{(t)} \}$ according to (\ref{eq52});

\STATE $t=t+1$.

\ENDWHILE

\RETURN $p_{m,n} = p_{m,n}^{(t)}$ for $\forall n,m$.

\end{algorithmic}

\end{algorithm}

Since (\ref{eq31}), $f(a)$ in (\ref{eq38}), and (\ref{eq46}) are
convex (or concave), the obtained $\{ c_{m,n}^{\left( t \right)}
\}$, $\{ a_{m,n}^{\left( t \right)} \}$, and $\{ p_{m,n}^{\left( t
\right)} \}$ are optimal solutions in the $t$th iteration.
Therefore, iteratively updating $\{ c_{m,n} \}$, $\{ a_{m,n} \}$,
and $\{ p_{m,n} \}$ will increase or maintain the value of the
objective function in (\ref{eq41})~\cite{QZRBF}. With the constraint
of the maximum transmitted power $P$, we will obtain a monotonically
non-decreasing sequence of the objective value in (\ref{eq41}) with
a upper bound, i.e., the global maximum. As a result, the proposed
iterative optimization algorithm for power allocation will converge to a stationary solution to the
problem (\ref{eq41}). To this end, we summarize the procedure of the proposed solution in
\textbf{Algorithm 1}.

In each iteration, the complexity to obtain the optimal $\{ c_{m,n} \}$ in (\ref{eq42}) and
$\{ a_{m,n} \}$ in (\ref{eq45}) is linear to the number of users, i.e., $\mathcal
{O} (K)$. $\lambda $ in (\ref{eq50}) and ${\mu _{m,n}} $ ($n = 1,2, \cdots
,{N_{\rm RF}}$, $m = 1,2, \cdots ,\left| {{S_n}} \right|$) in (\ref{eq51}) can be
obtained by using Newton¡¯s or bisection method with the complexity
$\mathcal {O} (K^2 {\log _2}\left( \delta \right))$, where
$\delta$ is the required accuracy. As a result, the complexity of the proposed power
allocation algorithm is $\mathcal {O} (T_{\max}K^2 {\log _2}\left( \delta  \right))$, where $T_{\max}$ is the maximum iteration times. Thus, the proposed iterative power allocation algorithm can be realized with a polynomial complexity.

\section{Simulation Results}\label{S6}

In this section, we provide the simulation results to verify the performance of the proposed beamspace MIMO-NOMA system. Specifically, we consider a typical downlink
mmWave massive MIMO system where the BS is equipped with an ULA of
$N=256$ antennas and communicates with $K$ users. The total
transmitted power is set as $P = 32$ mW (15 dBm)~\cite{XGBS}. One
LoS component and $L = 2$ NLoS components are assumed for all users'
channels. We consider the channel parameters of user $k$ as follows:
1) $\beta _k^{\left( 0 \right)} \sim \mathcal {CN}\left( {0,1}
\right)$, $\beta _k^{\left( l \right)} \sim \mathcal {CN}\left(
{0,{{10}^{ - 1}}} \right)$ for $1 \le l \le L$; 2) $\theta
_k^{\left( 0 \right)}$ and $\theta _k^{\left( l \right)}$ for $1 \le
l \le L$ follow the uniform distribution within $\left[ { -
\frac{1}{2},\frac{1}{2}} \right]$. The signal-to-noise ratio (SNR) is defined as
$\frac{E_b}{{{\sigma ^2}}}$ in this paper.

In the simulations, we consider the following four typical mmWave massive MIMO schemes for comparison: (1) ``Fully digital MIMO'', where each antenna is connected to one RF chain, i.e., $N_{\rm RF} = N$; (2) ``Beamspace MIMO''~\cite{XGBS}, where each beam only contains one user with $N_{\rm RF} = K$; (3) ``MIMO-OMA''~\cite{MSADPA} with $N_{\rm RF} \le K$, where OMA is performed for conflicting users, and users in the same beam are allocated with orthogonal frequency resources; (4) The ``proposed beamspace MIMO-NOMA'' with $N_{\rm RF} \le K$, which integrates NOMA and beamspace MIMO. Both the strongest user-based equivalent channel and SVD-based equivalent channel introduced in Section IV are considered, and the power allocation algorithm proposed in Section V is performed to alleviate interferences. Particularly, ZF precoding is considered in
fully digital MIMO and beamspace MIMO. In the following
subsections, the performance of the proposed beamspace MIMO-NOMA
scheme will be evaluated in terms of spectrum
efficiency\footnote{When the normalized bandwidth is considered, the
spectrum efficiency can defined as the achievable sum rate
(\ref{eq14}).} and energy efficiency.

\subsection{Spectrum Efficiency}

\begin{figure}[tp]
\begin{center}
\includegraphics[width=1\linewidth]{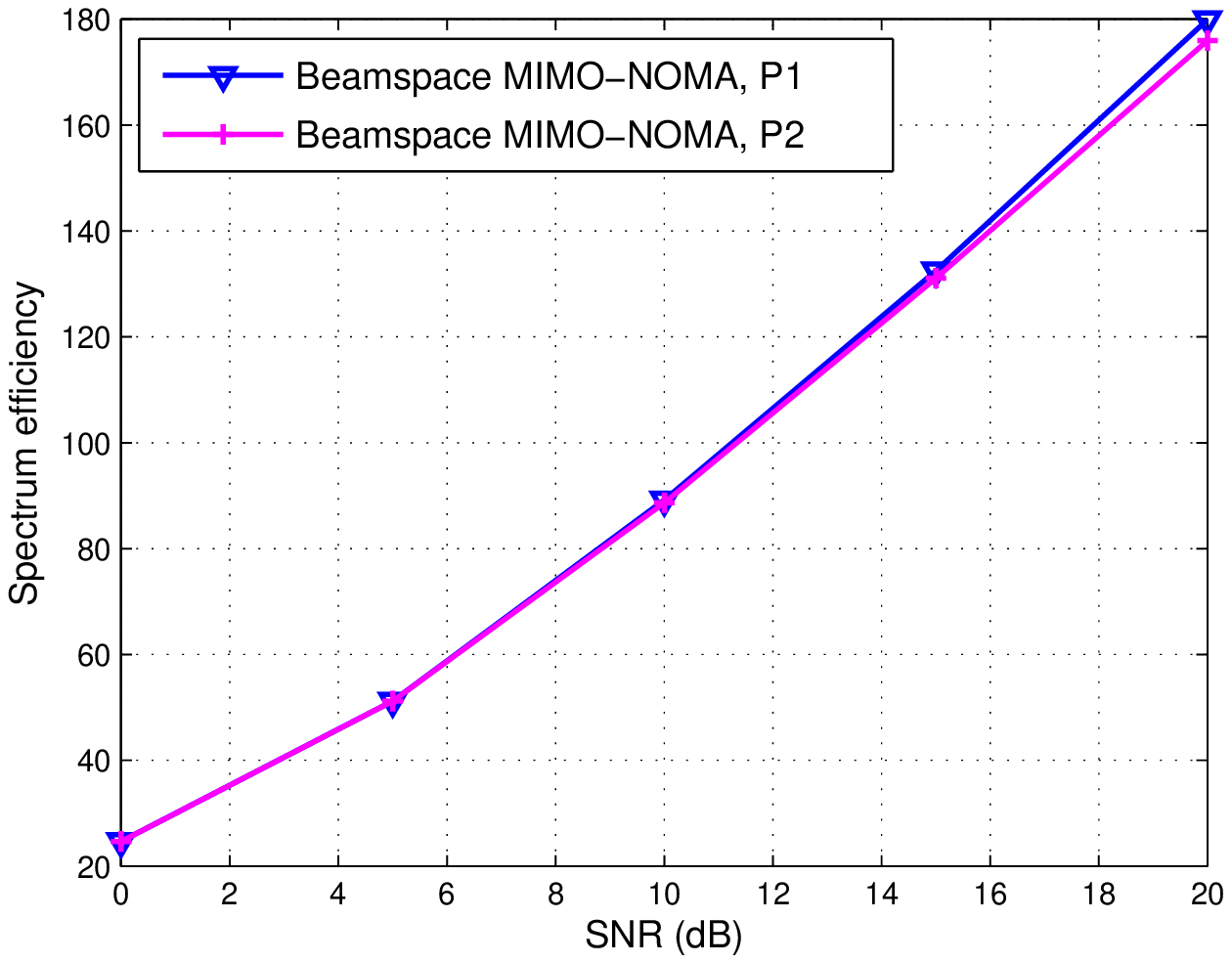} \caption{Spectrum efficiency against SNR, where the number of users is $K = 32$.}
\end{center}
\end{figure}

\begin{figure}[tp]
\begin{center}
\includegraphics[width=1\linewidth]{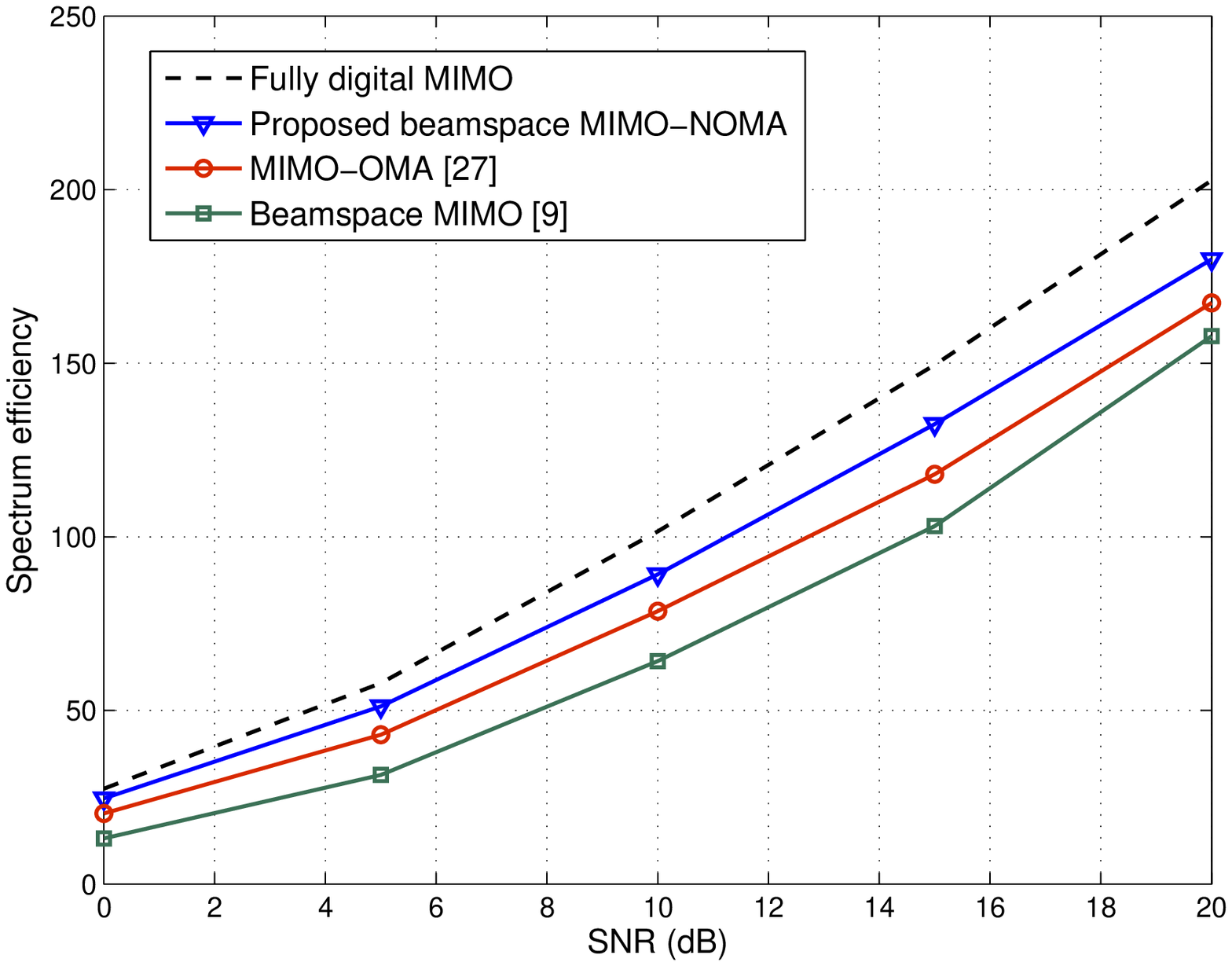} \caption{Spectrum efficiency against SNR, where the number of users is $K = 32$.}
\end{center}
\end{figure}

Fig. 2 shows the spectrum efficiency against SNR of the proposed
beamspace MIMO-NOMA scheme with the strongest user-based equivalent
channel (denoted as ``beamspace MIMO-NOMA, P1'') and SVD-based
equivalent channel (denoted as ``beamspace MIMO-NOMA, P2''), where
the number of users is $K = 32$ and the iteration times to solve the
power allocation optimization problem is set as 20, which is sufficient to make the iterative power allocation algorithm converged as shown later in Fig. 7. We can see from
the simulation results that ``beamspace MIMO-NOMA, P1'' with very low
complexity and ``beamspace MIMO-NOMA, P2'' with much higher
complexity caused by the SVD have very similar performance, which indicates that although
only the strongest user's channel is considered to perform precoding
in the proposed rank-deficient beamspace MIMO-NOMA scheme, it is
able to achieve the similar result compared to the precoding that
considers the effect of all users's channels in the same beam. This
favorable result is attributed to the strong correlation of beamspace
channels in the same beam. Thus, in the following simulations, we
only consider the strongest user-based equivalent channel to realize
low-complexity precoding since SVD is not required.

Fig. 3 shows the spectrum efficiency against SNR of the considered four schemes mentioned above, where the number of users is $K = 32$. We can find that the proposed beamspace MIMO-NOMA scheme can achieve higher spectrum efficiency than that of beamspace MIMO~\cite{XGBS} as well as MIMO-OMA~\cite{MSADPA}. Particularly, the proposed beamspace MIMO-NOMA has about 3 dB SNR gain compared to the beamspace MIMO, which benefits from the use of NOMA to serve multiple users in each beam. In addition, the proposed beamspace MIMO-NOMA also outperforms MIMO-OMA, since NOMA can achieve higher spectrum efficiency than that of OMA~\cite{KH2015nonsic,SMRNOMA,dai2015non}. It is intuitive that the fully digital MIMO can achieve the best spectrum efficiency as shown in Fig. 3, since there is not beam selection in the fully digital MIMO, and $N_{\rm RF}=N$ RF chains are used to serve all users. However, the fully digital MIMO suffers from the worst energy efficiency as shown in Fig. 5, which will be discussed later.

\begin{figure}[tp]
\begin{center}
\includegraphics[width=1\linewidth]{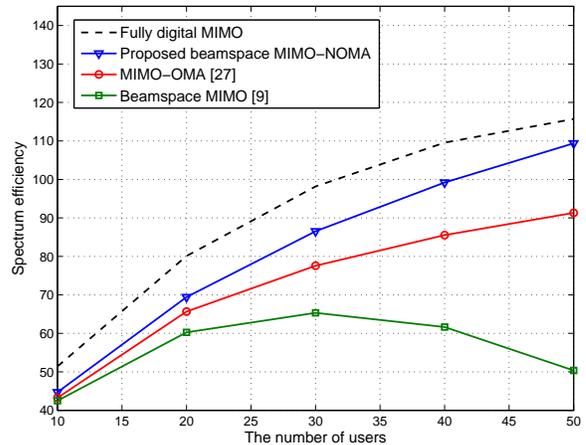} \caption{Spectrum efficiency against the number of users $K$, where SNR = 10 dB.}
\end{center}
\end{figure}

The performance comparison in terms of spectrum efficiency against
the number of users is shown in Fig. 4, where SNR is set as 10 dB.
We can see from the simulation results that with the increasing of
the number of users $K$, the performance gap between the
beamspace MIMO and the proposed beamspace MIMO-NOMA becomes larger.
This is because the larger the number of users, the larger the
probability that the same beam is selected for different users is.
As a result, existing beamspace MIMO will suffer from an obvious
performance loss, while the proposed beamspace MIMO-NOMA can still
perform well due to the use of NOMA.

\subsection{Energy Efficiency}

The energy efficiency $\varepsilon$ is defined as the ratio between the achievable sum rate $R_{\rm sum}$ and the total power consumption~\cite{XGEE}, i.e.,
\begin{equation}\label{eq54}
\varepsilon  = \frac{{{R_{\rm sum}}}}{{P + {N_{\rm RF}}{P_{\rm RF}}+
{N_{\rm RF}}{P_{\rm SW}}+P_{\rm BB}}} \ ({\rm bps/Hz/W}),
\end{equation}
where $P$ is the maximum transmitted power, $P_{\rm RF}$ is the power consumed by each RF chain, $P_{\rm SW}$ is the power consumption of switch, and $P_{\rm BB}$ is the baseband power consumption. Particularly, we adopt
the typical values $P_{\rm RF} = 300$ mW, $P_{\rm SW} = 5$ mW, and $P_{\rm BB} = 200$ mW~\cite{XGEE}.

\begin{figure}[tp]
\begin{center}
\includegraphics[width=1\linewidth]{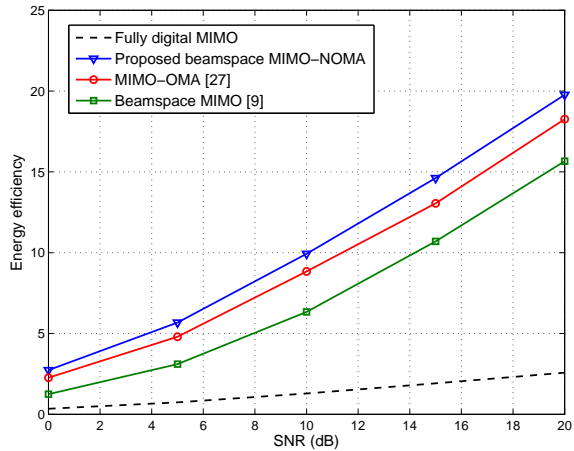} \caption{Energy efficiency against SNR, where the number of users is $K = 32$.}
\end{center}
\end{figure}

\begin{figure}[tp]
\begin{center}
\includegraphics[width=1\linewidth]{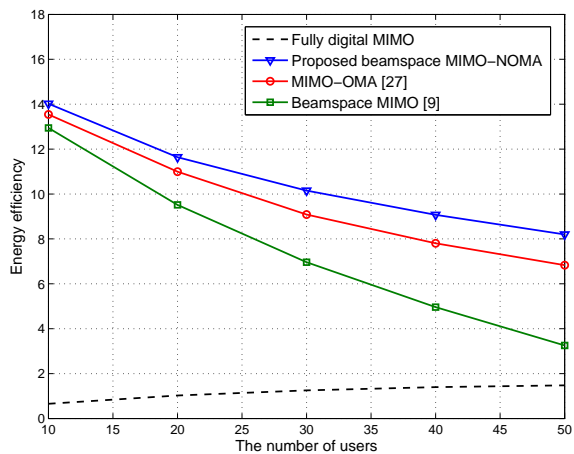} \caption{Energy efficiency against the number of users $K$, where SNR = 10 dB.}
\end{center}
\end{figure}

Fig. 5 shows the energy efficiency against SNR, where the number of users is also $K = 32$. We can find that the proposed beamspace MIMO-MOMA can achieve higher energy efficiency than other three schemes. Particularly, the proposed beamspace MIMO-MOMA has about $25 \%$ energy efficiency improvement compared to existing beamspace MIMO, which benefits from the use of NOMA to serve multiple users in each beam. In addition, the proposed beamspace MIMO-MOMA can achieve much higher energy efficiency than the fully digital MIMO scheme, where the number of RF chains is equal to the number of BS antennas, which leads to very high energy consumption, e.g., 300 mW for each RF chain. On the contrary, the number of RF chains is much smaller than the number of antennas in the proposed beamspace MIMO-MOMA scheme. Therefore, the energy consumption caused by the RF chains can be significantly reduced compared to the fully digital MIMO scheme.

The performance comparison in terms of energy efficiency against the
number of users is shown in Fig. 6, where SNR is set as 10 dB. We
can see that the energy efficiency of
the proposed beamspace MIMO-MOMA scheme is higher than all of other
three schemes even the number of users is very large (e.g., 50 users
are simultaneously severed).

\subsection{Convergence of power allocation}

\begin{figure}[tp]
\begin{center}
\includegraphics[width=1\linewidth]{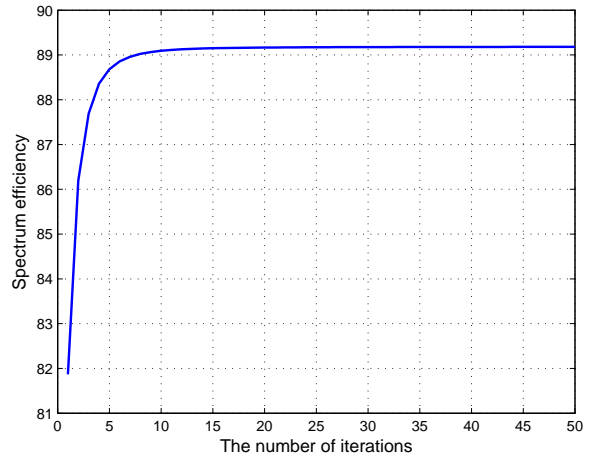} \caption{Spectrum efficiency against the number of iterations for power allocation.}
\end{center}
\end{figure}

In this subsection, we evaluate the convergence of the proposed
iterative power allocation algorithm in Section V, where the number of
users is set as $K = 32$, and SNR = 10 dB. As shown in Fig. 7, the
spectrum efficiency tends to be stable after 10 times of iteration,
which verifies the convergence of the proposed power allocation as
discussed in Section V.

\subsection{The user fairness}

\begin{figure}[tp]
\begin{center}
\includegraphics[width=1\linewidth]{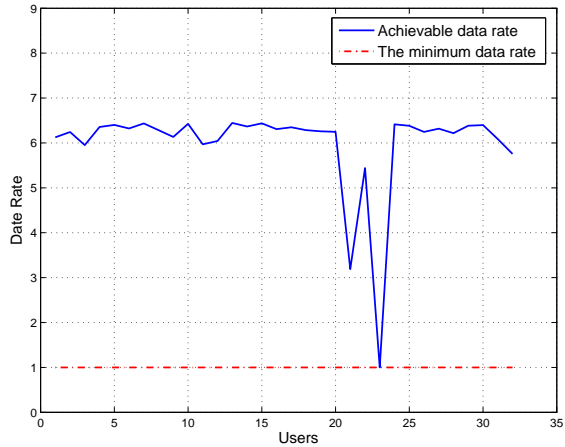} \caption{The achievable data rate for each user.}
\end{center}
\end{figure}

In this subsection, we evaluate the user fairness of the proposed beamspace MIMO-NOMA scheme, where the number of users is set as $K = 32$, SNR = 20 dB, and $R_{\rm min}=1$ bit/s/HZ. As shown in Fig. 8, the achievable data rate for each user is larger than the minimum data rate $R_{\rm min}$, due to the data rate constraint for each user, i.e., $C_3$ in (23).

\section{Conclusions}\label{S7}

In this paper, we have proposed a new transmission scheme, i.e., beamspace
MIMO-NOMA, to integrate NOMA and beamspace MIMO to break the fundamental limit of existing beamspace MIMO that only one user can be served in each beam at the same time-frequency resources.
Particularly, the number of users can be larger than the number of
RF chains in the proposed beamspace MIMO-NOMA scheme, which is essentially different from existing
beamspace MIMO systems. In addition to realizing
massive connectivity, the use of NOMA can also improve the capacity
bound of beamspace MIMO. To restrain the inter-beam
interferences, the equivalent channel vector is determined for each
beam to realize precoding based on the principle of ZF, which
considers the high correlation of users' beamspace channels in the same beam at
mmWave frequencies. Furthermore, to suppress both
inter-beam and intra-beam interferences, we proposed to jointly
optimize the power allocation of all users by maximizing the
achievable sum rate, and an iterative optimization algorithm has
been developed to realize power allocation. Simulation results have
shown that the proposed beamspace MIMO-NOMA can achieve better
performance in terms of spectrum and energy efficiency compared to
existing beamspace MIMO, e.g., $25 \%$ energy efficiency gain
can be achieved. In the future, we
will consider sophisticated user pairing/clustering for the proposed
beamspace MIMO-NOMA scheme in ultra-dense network (UDN).

\begin{IEEEbiography}[{\includegraphics[width=0.9in,height=1.1in,keepaspectratio]{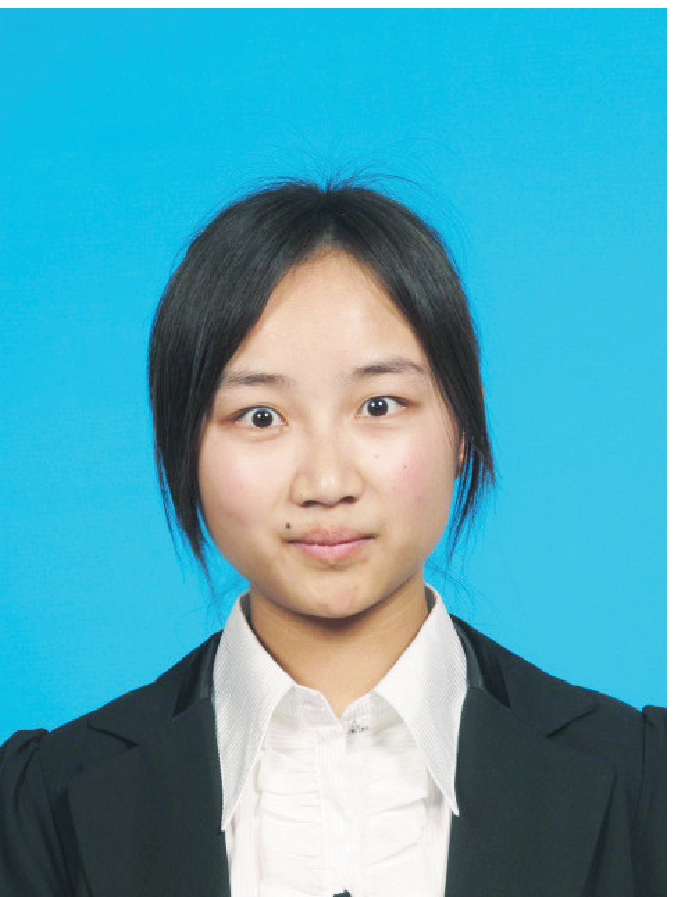}}]{Bichai Wang}
(S'15) received her B.S. degree in Electronic Engineering from
Tsinghua University, Beijing, China, in 2015. She is currently
working towards the Ph.D. degree in the Department of Electronic
Engineering, Tsinghua University, Beijing, China. Her research
interests are in wireless communications, with the emphasis on new
multiple access techniques. She has received the Freshman
Scholarship of Tsinghua University in 2011, the Academic Merit
Scholarships of Tsinghua University in 2012, 2013, and 2014,
respectively, the Excellent Thesis Award of Tsinghua University
in 2015, and the National Scholarship in 2016.
\end{IEEEbiography}

\begin{IEEEbiography}[{\includegraphics[width=0.9in,height=1.1in,keepaspectratio]{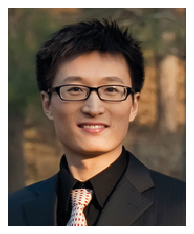}}]{Linglong Dai}
(M'11-SM'14) received the B.S. degree from Zhejiang University in 2003, the M.S. degree (with the highest honor) from the China Academy of Telecommunications Technology (CATT) in 2006, and the Ph.D. degree (with the highest honor) from Tsinghua University, Beijing, China, in 2011. From 2011 to 2013, he was a Postdoctoral Research Fellow with the Department of Electronic Engineering, Tsinghua University, where he has been an Assistant Professor since July 2013 and then an Associate Professor since June 2016. His current research interests include massive MIMO, millimeter-wave communications, multiple access, and sparse signal processing. He has published over 50 IEEE journal papers and over 30 IEEE conference papers. He also holds 13 granted patents. He has received 4 conference Best Paper Awards at IEEE ICC 2013, IEEE ICC 2014, WCSP 2016, and IEEE ICC 2017, and he also received the IEEE Transactions on Broadcasting Best Paper Award in 2015. He currently serves as Editor of IEEE Transactions on Communications, IEEE Transactions on Vehicular Technology, and IEEE Communications Letters.
\end{IEEEbiography}

\begin{IEEEbiography}[{\includegraphics[width=1.0in,height=1.2in,clip,keepaspectratio]{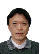}}]{Zhaocheng Wang}
(M'09-SM'11) received his B.S., M.S., and Ph.D. degrees from Tsinghua University, Beijing, China, in 1991, 1993, and 1996, respectively.
From 1996 to 1997, he was a Post-Doctoral Fellow with Nanyang Technological University, Singapore. From 1997 to 1999, he was with the OKI Techno
Centre (Singapore) Pte. Ltd., Singapore, where he was a Research Engineer and then became a Senior Engineer. From 1999 to 2009, he was with Sony
Deutschland GmbH, where he was a Senior Engineer and then became a Principal Engineer. He is currently a Professor of Department of Electronic Engineering
with Tsinghua University and serves as the Director of the Broadband Communication Key Laboratory, Tsinghua National Laboratory
for Information Science and Technology. His research interests include wireless communications, visible light communications, millimeterwave communications,
and digital broadcasting. He has authored or co-authored over 130 journal papers and holds 34 granted U.S./EU patents. He has co-authored two books,
one of which, Millimeter Wave Communication Systems, was selected by the IEEE Series on Digital and Mobile Communication (Wiley-IEEE Press).
He received 2013 Beijing Science and Technology Award (First Prize), IEEE ICC 2013 Best Paper Award, OECC 2015 Best Student Award,
2016 IEEE Scott Helt Memorial Award (Best Paper Award of IEEE Transactions on Broadcasting), 2016 National Award for Science and Technology Progress (First Prize)
and IEEE ICC 2017 Best Paper Award. He is a Fellow of the Institution of Engineering and Technology. He served as an Associate Editor of the IEEE Transactions on
Wireless Communications from 2011 to 2015 and an Associate Editor of the IEEE Communications Letters from 2013 to 2016, and has also served as the
technical program committee co-chairs of various international conferences.
\end{IEEEbiography}

\begin{IEEEbiography}[{\includegraphics[width=1.0in,height=1.2in,clip,keepaspectratio]{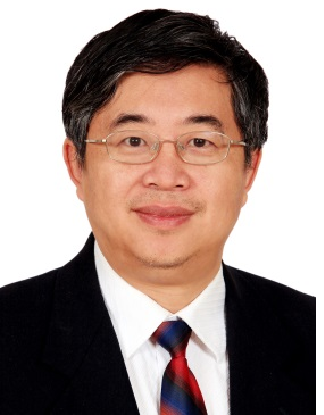}}]{Ning Ge}
received his B.S. degree in 1993, and his Ph.D. in 1997, both from Tsinghua University, China. From 1998 to 2000, he worked on the development of ATM switch fabric ASIC in ADC Telecommunications, Dallas. Since 2000, he has been with the Department of Electronics Engineering at Tsinghua University, where he is a professor and serves as Director of Communication Institute. His research interests include ASIC design, short range wireless communication, and wireless communications. He is a senior member of CIC and CIE, and a member of the IEEE.
\end{IEEEbiography}

\begin{IEEEbiography}[{\includegraphics[width=1.0in,height=1.2in,clip,keepaspectratio]{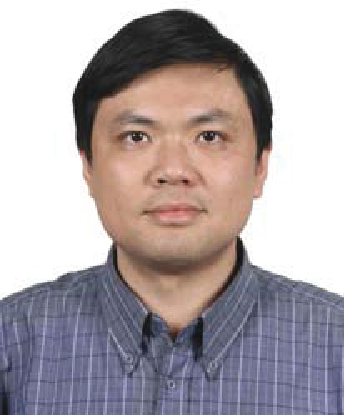}}]{Shidong Zhou}
(M'98) received B.S. and M.S. degrees from Southeast University, Nanjing, China, in 1991 and 1994, respectively, and the Ph.D. degree from Tsinghua University, Beijing, China, in 1998. He is now a Professor with the Department of Electronic Engineering, Tsinghua University. He was engaged in several major national projects on 3G and 4G mobile communication technique research and development. His research interests include mobile communication system architectures, advanced transmission technique, wireless channel sounding and modeling, radio resource management, and high energy efficient wireless networks.
\end{IEEEbiography}


\begin{thebibliography}{}
\providecommand{\url}[1]{#1}
\csname url@samestyle\endcsname
\providecommand{\newblock}{\relax}
\providecommand{\bibinfo}[2]{#2}
\providecommand{\BIBentrySTDinterwordspacing}{\spaceskip=0pt\relax}
\providecommand{\BIBentryALTinterwordstretchfactor}{4}
\providecommand{\BIBentryALTinterwordspacing}{\spaceskip=\fontdimen2\font plus
\BIBentryALTinterwordstretchfactor\fontdimen3\font minus
  \fontdimen4\font\relax}
\providecommand{\BIBforeignlanguage}[2]{{%
\expandafter\ifx\csname l@#1\endcsname\relax
\typeout{** WARNING: IEEEtran.bst: No hyphenation pattern has been}%
\typeout{** loaded for the language `#1'. Using the pattern for}%
\typeout{** the default language instead.}%
\else
\language=\csname l@#1\endcsname
\fi
#2}}
\providecommand{\BIBdecl}{\relax}
\BIBdecl

\end{thebibliography}


\begin{thebibliography}{1}


\bibitem{MMWAVE}
S. Mumtaz, J. Rodriquez, and L. Dai, \emph{MmWave Massive MIMO: A Paradigm for 5G}, Academic Press, Elsevier, 2016.

\bibitem{JBBS}
J. Brady, N. Behdad, and A. Sayeed, ``Beamspace MIMO for millimeterwave
communications: System architecture, modeling, analysis, and
measurements,'' \emph{IEEE Trans. Antennas Propag.}, vol. 61, no. 7, pp. 3814-
3827, Jul. 2013.

\bibitem{HX2}
H. Xie, B. Wang, F. Gao, and S. Jin, ``A full-space spectrum-sharing strategy for massive MIMO cognitive radio,'' \emph{IEEE J. Select. Areas Commun.}, vol. 34, no. 10, pp. 2537-2549, Oct. 2016.

\bibitem{ASBSMIMO}
A. Sayeed and J. Brady, ``Beamspace MIMO for high-dimensional
multiuser communication at millimeter-wave frequencies,'' in
\emph{Proc. IEEE Global Communications Conference (IEEE
GLOBECOM'13)}, Dec. 2013, pp. 3679-3684.

\bibitem{PARF}
P. Amadori and C. Masouros, ``Low RF-complexity millimeter-wave
beamspace-MIMO systems by beam selection,'' \emph{IEEE Trans. Commun.},
vol. 63, no. 6, pp. 2212-2222, Jun. 2015.

\bibitem{XGEE}
X. Gao, L. Dai, S. Han, C.-L. I, and R. W. Heath, ``Energy-efficient hybrid analog and digital precoding for mmWave MIMO systems with large antenna arrays,'' \emph{IEEE J. Sel. Areas Commun.}, vol. 34, no. 4, pp. 998-1009, Apr. 2016.

\bibitem{HX1}
H. Xie, F. Gao, S. Zhang, and S. Jin, ``A unified transmission strategy for TDD/FDD massive MIMO systems with spatial basis expansion model,'' \emph{IEEE Trans. Veh. Technol.}, vol. 66, no. 4, pp. 3170-3184, Apr. 2017.

\bibitem{XC1}
X. Cheng, B. Yu, L. Yang, J. Zhang, G. Liu, Y. Wu, and L. Wan, ``Communicating in the real world: 3D MIMO,'' \emph{IEEE Wireless Commun. Mag.}, vol. 21, no. 4, pp. 136-144, Aug. 2014.

\bibitem{XGBS}
X. Gao, L. Dai, Z. Chen, Z. Wang, and Z. Zhang, ``Near-optimal beam selection for beamspace mmWave massive MIMO systems,'' \emph{IEEE Commun. Lett.}, vol. 20, no. 5, pp. 1054-1057, May 2016.

\bibitem{FRMIMO}
F. Rusek et al., ``Scaling up MIMO: Opportunities and challenges with
very large arrays,'' \emph{IEEE Signal Process. Mag.}, vol. 30, no. 1, pp. 40-60,
Jan. 2013.

\bibitem{SCEE}
S. Cui, A. J. Goldsmith, and A. Bahai, ``Energy-efficiency of MIMO and cooperative MIMO techniques in sensor networks,'' \emph{IEEE J. Sel. Areas Commun.}, vol. 22, no. 6, pp. 1089-1098, Aug. 2004.

\bibitem{SSAS}
S. Sanayei and A. Nosratinia, ``Antenna selection in MIMO systems,'' \emph{IEEE Commun. Mag.}, vol. 42, no. 10, pp. 68-73, Oct. 2004.

\bibitem{AMAS}
A. Molisch and M. Win, ``MIMO systems with antenna selection,'' \emph{IEEE Microw. Mag.}, vol. 5, no. 1, pp. 46-56, Mar. 2004.

\bibitem{SSCAS}
S. Sanayei and A. Nosratinia, ``Capacity of MIMO channels with antenna selection,'' \emph{IEEE Trans. Inf. Theory}, vol. 53, no. 11, pp. 4356-4362, Nov. 2007.

\bibitem{YZMM}
Y. Zeng and R. Zhang, ``Millimeter wave MIMO with lens antenna array:
A new path division multiplexing paradigm,'' \emph{IEEE Trans. Commun.}, vol. 64, no. 4, pp. 1557-1571, Apr. 2016.

\bibitem{ZD1}
Z. Ding, Z. Yang, P. Fan, and  H. V. Poor, ``On the performance of non-orthogonal multiple access in 5G systems with randomly deployed users,'' \emph{IEEE Signal Process. Lett.}, vol. 21, no. 12, pp. 1501-1505, Dec. 2014.

\bibitem{KH2015nonsic}
K. Higuchi and A. Benjebbour, ``Non-orthogonal multiple access
(NOMA) with successive interference cancellation for future radio
access,'' \emph{IEICE Trans. Commun.}, vol. E98-B, no. 3, pp. 403-414,
Mar. 2015.

\bibitem{ZD2}
Z. Ding, P. Fan, and H. V. Poor, ``Impact of user pairing on 5G non-orthogonal multiple-access downlink transmissions,'' \emph{IEEE Trans. Veh. Technol.}, vol. 65, no. 8, pp. 6010-6023, Aug. 2016.

\bibitem{SMRNOMA}
S. M. R. Islam, N. Avazov, O. A. Dobre, and K.-S. Kwak,
``Power-domain non-orthogonal multiple access (NOMA) in 5G systems: Potentials and challenges,'' to appear in \emph{IEEE Commun. Surveys \& Tutorials}, 2017.

\bibitem{ZDAMIMO}
Z. Ding, F. Adachi, and H. V. Poor, ``The application of MIMO to
non-orthogonal multiple access,'' \emph{IEEE Trans. Wireless
Commun.}, vol. 15, no. 1, pp. 537-552, Jan. 2016.

\bibitem{dai2015non}
L.~Dai, B.~Wang, Y.~Yuan, S.~Han, C.-L. I, and Z.~Wang,
``Non-orthogonal multiple access for 5G: Solutions, challenges, opportunities, and future research trends,'' \emph{IEEE Commun. Mag.}, vol. 53, no. 9, pp. 74-81, Sep. 2015.

\bibitem{ZDLDMIMONOMA}
Z. Ding, L. Dai, and H. V. Poor, ``MIMO-NOMA design for small packet
transmission in the Internet of Things,'' \emph{IEEE Access}, vol.
4, pp. 1393-1405, Aug. 2016.

\bibitem{ZDMMIMO}
Z. Ding and H. V. Poor, ``Design of massive-MIMO-NOMA with limited
feedback,'' \emph{IEEE Signal Process. Lett.}, vol. 23, no. 5, pp.
629-633, May 2016.

\bibitem{YSNOMA}
Y. Saito, Y. Kishiyama, A. Benjebbour, and T. Nakamura,
``Non-orthogonal multiple access (NOMA) for cellular future radio
access,'' in \emph{Proc. IEEE Vehicular Technology
  Conference (IEEE VTC Spring'13)}, Jun. 2013, pp. 1-5.

\bibitem{YHPA}
Y. Hayashi, Y. Kishiyama, and K. Higuchi, ``Investigations on power
allocation among beams in non-orthogonal access with random
beamforming and intra-beam SIC for cellular MIMO downlink,'' in
\emph{Proc. IEEE Vehicular Technology
  Conference (IEEE VTC Fall'13)}, Sep. 2013, pp. 1-5.

\bibitem{BKNOMABF}
B. Kim, S. Lim, H. Kim, S. Suh, J. Kwun, S. Choi, C Lee, S. Lee,
and D. Hong, ``Non-orthogonal multiple access in a downlink multiuser
beamforming system,'' in \emph{Proc. IEEE Military Communications Conference (IEEE MILCOM'13)}, Nov. 2013,
pp. 1278-1283.

\bibitem{MSADPA}
M. S. Ali, H. Tabassum, and E. Hossain, ``Dynamic user clustering and
power allocation in non-orthogonal multiple access (NOMA) systems,''
\emph{IEEE Access}, vol. 4, pp. 6325-6343, Aug. 2016.

\bibitem{[25]}
M. S. Ali, E. Hossain, and D. I. Kim, ``Non-orthogonal multiple access (NOMA) for downlink multiuser MIMO systems: User clustering, beamforming, and power allocation,'' \textit{IEEE Access}, vol. 5, pp. 565-577, Dec. 2016.

\bibitem{QSNOMA}
Q. Sun, S. Han, Chin-Lin I, and Z. Pan, ``On the ergodic capacity of
MIMO NOMA systems,'' \emph{IEEE Wireless Commun. Lett.}, vol. 4, no. 4, pp. 405-408,
Aug. 2015.

\bibitem{QZRBF}
Q. Zhang, Q. Li, and J. Qin, ``Robust beamforming for non-orthogonal
multiple access systems in MISO channels,'' to appear in \emph{IEEE Trans. Veh.
Technol.}, 2017.

\bibitem{[28]}
M. F. Hanif, Z. Ding, T. Ratnarajah, and G. K. Karagiannidis ``A minorization-maximization method for optimizing sum rate in non-orthogonal multiple access systems'', \textit{IEEE Trans. Signal Process.}, vol. 64, no. 1, pp. 76-88, Jan. 2016.

\bibitem{[29]}
Z. Ding, P. Fan, and H. V. Poor, ``Random beamforming in millimeter-wave NOMA networks,'' to appear in \textit{IEEE Access}, 2017.

\bibitem{TSRICC}
T. S. Rappaport, E. Ben-Dor, J. N. Murdock, and Y. Qiao, ``38 GHz and 60 GHz angle-dependent propagation for cellular \& Peer-to-Peer Wireless Communications,'' in \emph{Proc. IEEE International Conference on Communications (ICC'12)}, Jun. 2012, pp. 4568-4573.

\bibitem{SHLS}
S. Han, C.-L. I, Z. Xu, and C. Rowell, ``Large-scale antenna systems with
hybrid precoding analog and digital beamforming for millimeter wave
5G,'' \emph{IEEE Commun. Mag.}, vol. 53, no. 1, pp. 186-194, Jan. 2015.

\bibitem{QHZF}
Q. H. Spencer, A. L. Swindlehurst, and M. Haardt, ``Zero-forcing
methods for downlink spatial multiplexing in multiuser MIMO
channels,'' \emph{IEEE Trans. Signal Process.}, vol. 52, no. 2, pp.
461-471, Feb. 2004.

\bibitem{JRMDC}
J. R. Magnus and H. Neudecher, \emph{Matrix Differential Calculus with Application in Statistics and Econometrics}. New York, NY, USA: Wiley, 1988.

\bibitem{SBCV}
S. Boyd and L. Vandenberghe, \emph{Convex Optimication}. Cambridge,
U. K.: Cambridge Univ. Press, 2004.

\end{thebibliography}
\end{document}